\documentclass[aps,physrev,reprint,superscriptaddress]{revtex4-1}
\usepackage{graphicx}% Include figure files
\usepackage{dcolumn}% Align table columns on decimal point
\usepackage{bm}% bold math
\usepackage{hyperref}% add hypertext capabilities
\usepackage{float}
\usepackage{booktabs} % in preamble
\usepackage{multirow}

\usepackage{comment}
\usepackage[english]{babel}
\begin{document}

%-----Title------%
%\title{Modeling Plant Disease Epidemics in Networks}
\title{Modeling plant disease spread via high-resolution human mobility networks}
\author{Varun K. Rao}
\affiliation{Center for Complex Networks and Systems Research, Luddy School of Informatics, Computing, and Engineering, Indiana University, Bloomington, Indiana 47408, USA}
\affiliation{School of Public Health, Indiana University, Bloomington, Indiana 47408, USA}
\author{Ryan Higgs}
\affiliation{Onside, Christchurch,  New Zealand.}
\author{Hautahi Kingi}
\affiliation{Onside, Christchurch,  New Zealand.}
\author{Filippo Radicchi}
\affiliation{Center for Complex Networks and Systems Research, Luddy School of Informatics, Computing, and Engineering, Indiana University, Bloomington, Indiana 47408, USA}
\author{Santo Fortunato}
\affiliation{Center for Complex Networks and Systems Research, Luddy School of Informatics, Computing, and Engineering, Indiana University, Bloomington, Indiana 47408, USA}
\author{Maria Litvinova}
\email{malitv@iu.edu}
\affiliation{School of Public Health, Indiana University, Bloomington, Indiana 47408, USA}
%TC:incbib

% abstract is 194 words 
\begin{abstract}
Human mobility plays a crucial role in the spread of human diseases, but is rarely quantified in plant disease epidemics. 
To address this gap, we integrate a unique, high-resolution network of human movements in New Zealand with a metapopulation model to mechanistically simulate pathogen transmission. 
We calibrate the model on the nationwide 2010 kiwifruit vine disease (Psa-V) outbreak, and show that it accurately reproduces the observed spatiotemporal spread, confirming that the human mobility network is a strong foundation for modeling transmission dynamics. By analyzing spatial infection trends, we find that most dispersal occurs locally, as often illustrated in the plant-outbreak literature. However, sporadic long-range connections are necessary to model a nationwide outbreak.
Using the model as an in-silico laboratory, we demonstrate that enhanced surveillance accelerates detection and that outbreak severity is highly sensitive to the timing and location of initial disease importation. We observe a potential causal link between seasonal labor patterns and epidemic risk in high-traffic seasons. 
This study provides a robust, data-driven framework for modeling and predicting the spatiotemporal spread of agricultural pathogens. It underscores the importance of leveraging human mobility networks to design timely interventions and surveillance systems, protecting global food security. 
\end{abstract}

%Use the showkeys class option if keyword

 %display desired
\maketitle
%\tableofcontents

\section{Introduction}

% NEW IDEA: plant disease spreads in many ways, and is modeled in very specific ways

Direct and indirect impacts of human infections on population health and economic well-being captured the attention of scientific and general audiences in recent years, overshadowing the impact of plant disease outbreaks on national and international biosecurity. Based on a review conducted by Fielder \textit{et al.}~\cite{fielder2024synoptic}, at least 38 new plant pathogens were identified, while 10 global plant outbreaks in 2023 challenged crop production, food prices, and food security. 
  Predicting the dynamics of plant disease outbreaks, and consequently responding to these persistent threats, is becoming increasingly complex~\cite{laine2023plant} due to changes in pathogens, their hosts, climate, and human behaviors. Because plant pathogens spread via wind, rain, and insects, studies have emphasized environment-related landscape modeling studies that simulate host-pathogen interactions. These landscape-focused studies use a variety of approaches, including lattice-based models using kernel functions for dispersal~\cite{yuen2015landscape, cendoya_individual-based_2024,orozco2019early,landry_spatio-temporal_2021,signes-pont_epidemic_2020,white2017modelling,severns2022dispersal,chapman2025modelling,suprunenko2025predicting}, weather-related statistical modeling~\cite{shah2019predicting,gent2019prediction}, Bayesian methods~\cite{adrakey2023bayesian,varghese2020estimating}, meta-population modeling~\cite{severns2022dispersal,rimbaud2018using}, and 
mean-field approximations~\cite{forster2007optimizing} to understand and predict the spread of these pathogens.  However, with growing connectivity, human-mediated spread through contaminated equipment and transport~\cite{ocimati2021farmer, hardy_pathway_2013,hodkinson1997plant}, infected plant material~\cite{ristaino2000new}, or global trade~\cite{pautasso2014network} has become an important force in driving rapid, medium- and long-distance transmission events. 

% GAP in modeling techniques and how we ADDRESS GAP generally,
Nevertheless, there are few studies investigating and modeling the impact of human mobility~\cite{marshall2021assessing,andersen2019modeling,moslonka2011networks,benedetti_spatial_2017,numminen2020spread} on the long-distance spread of plant-related pathogens~\cite{jeger_modelling_2007,pautasso2014network}. Other works studying long-distance spread  have focused on weather-related spread \cite{aylor2003spread,brown2002aerial}. The few existing studies concerning the role of between-property mobility networks focus on trading~\cite{marshall2021assessing,andersen2019modeling,moslonka2011networks}, transport networks~\cite{benedetti_spatial_2017,numminen2020spread}, or geographical distance~\cite{sutrave2012identifying,strona2017network}. However, these studies use static networks to model spread, and do not account for the temporal changes in the magnitude of the traffic between the properties. Using inspiration from studies modeling spreading processes on temporal networks of livestock movements we represent snapshots of movements at different time periods~\cite{bajardi2011dynamical,bajardi2012optimizing,valdano2015predicting,schirdewahn2021early} to fill this gap. We propose a multi-scale meta-population model based on empirical network data of human mobility between agricultural properties. Our approach combines lattice model for within-property spread and create a mobility-based network model for between-property spread.

 %and where DATA comes from
 Our work eliminates the main barrier to integrating human mobility into plant disease epidemiology: the lack of high-quality spatiotemporal data suitable for predictive modeling. Our unique dataset for the horticulture industry of New Zealand provides time-stamped origin-destination data for movements between properties. The data was provided and managed by Onside, an agritech company with biosecurity technology covering more than 20,000 properties from multiple industries across several countries, including New Zealand and Australia. 

% talk about inspiration from animal literature and give overview of findings in paper 

We leverage a Susceptible-Infected metapopulation model~\cite{colizza2008epidemic} to describe plant disease dynamics and calibrate the model to replicate the timing and severity of a past outbreak of \textit{Pseudomonas syringae pv. actinidiae} (Psa-V) \cite{birnie2014lessons}; a bacterial disease that was introduced to New Zealand in 2010 and is still present and widespread throughout the country. The disease devastated the country's kiwifruit industry~\cite{birnie2014lessons,vanneste2017scientific,froud2014relationships}, and was projected to cost kiwifruit growers $\$310$ to $\$410$ million dollars for five years after the outbreak \cite{greer2012costs} due to a combination of fast spread and pathogen severity ~\cite{froud2015review,cameron2014pseudomonas}. Human-mediated transmission, caused by movement of infected materials between properties \cite{hardy_pathway_2013}, contributes to this rapid spread since the bacteria can stay alive on machinery or tools \cite{froud2015review}. Using this fact, our data-driven network-based spreading model is calibrated to this outbreak's severity.  We explore three different temporal aggregations of the network, similar to Bajardi \textit{et al.} \cite{bajardi2011dynamical}, to reflect horticulture industry seasonality. By using this calibrated model on seasonal networks, we illustrate that the month of pathogen importation strongly dictates epidemic size, peak, and onset time, demonstrating a direct link between movement seasonality and outbreak dynamics. This provides a practical guide to the development, calibration, and utilization of computational models that take advantage of human mobility data for predictive plant disease epidemiology, targeted surveillance, and biosecurity.

\section{Methods}
% Go over general approach here . . 
\subsection{Movement Data}
 % NZ MAP FIGURE
\begin{figure}
    \centering
    \includegraphics[width=1.2\linewidth]{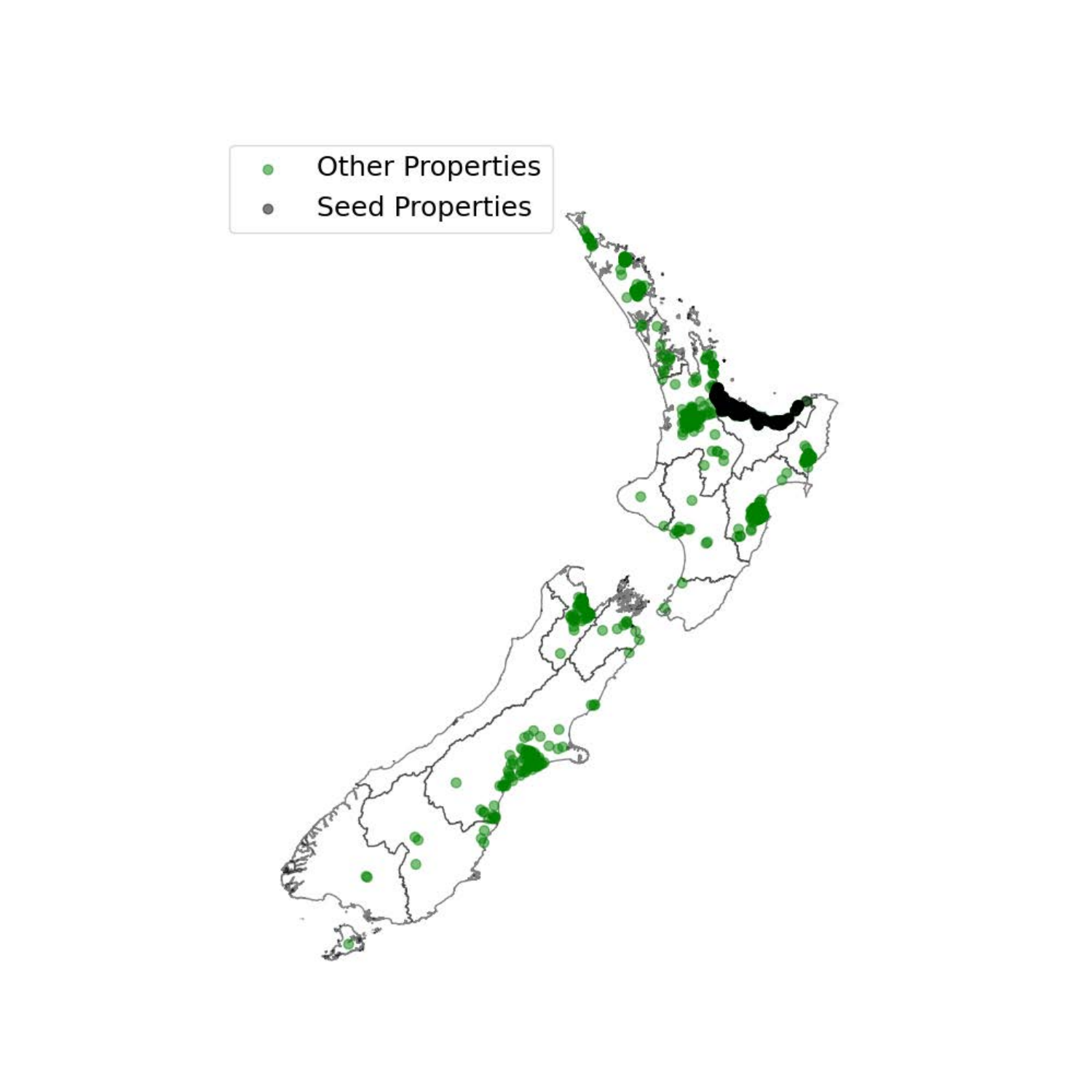}
    \caption{\textbf{Map of the horticulture industry in New Zealand.} Each point represents a property in the Onside dataset ($N=2,281$ total properties). Black points denote the $S=1,300$ properties located in the Bay of Plenty, which are used as seeds for the spreading dynamics simulated by our model. }
    \label{fig:nz-seeds}
\end{figure}

% Discuss data source

We rely on data recorded by the Onside mobile app. The app is used by farmers and businesses servicing rural properties, such as contractors, processors, and other service providers. Data includes the information when users enter (``check-in'') one of the registered properties. Check-in events can either be recorded manually (i.e., a user checks in arriving at a property) or automatically (i.e., triggered by reaching the property boundaries based on digital geofence technology) and are associated with time stamps having a resolution of minutes.

Onside data contains information about individual properties, such as their area and geographical coordinates, and are classified based on their industry: horticulture, agriculture, and farming. Since we aim to model the Psa-V outbreak in New Zealand, we only focus on the $N = 2,281$ properties in the horticulture industry. A geographical map of the horticultural system in New Zealand reconstructed using Onside data is visualized in Figure \ref{fig:nz-seeds}.

\begin{figure*}[!htb]
    \centering
    \includegraphics[width=\linewidth]{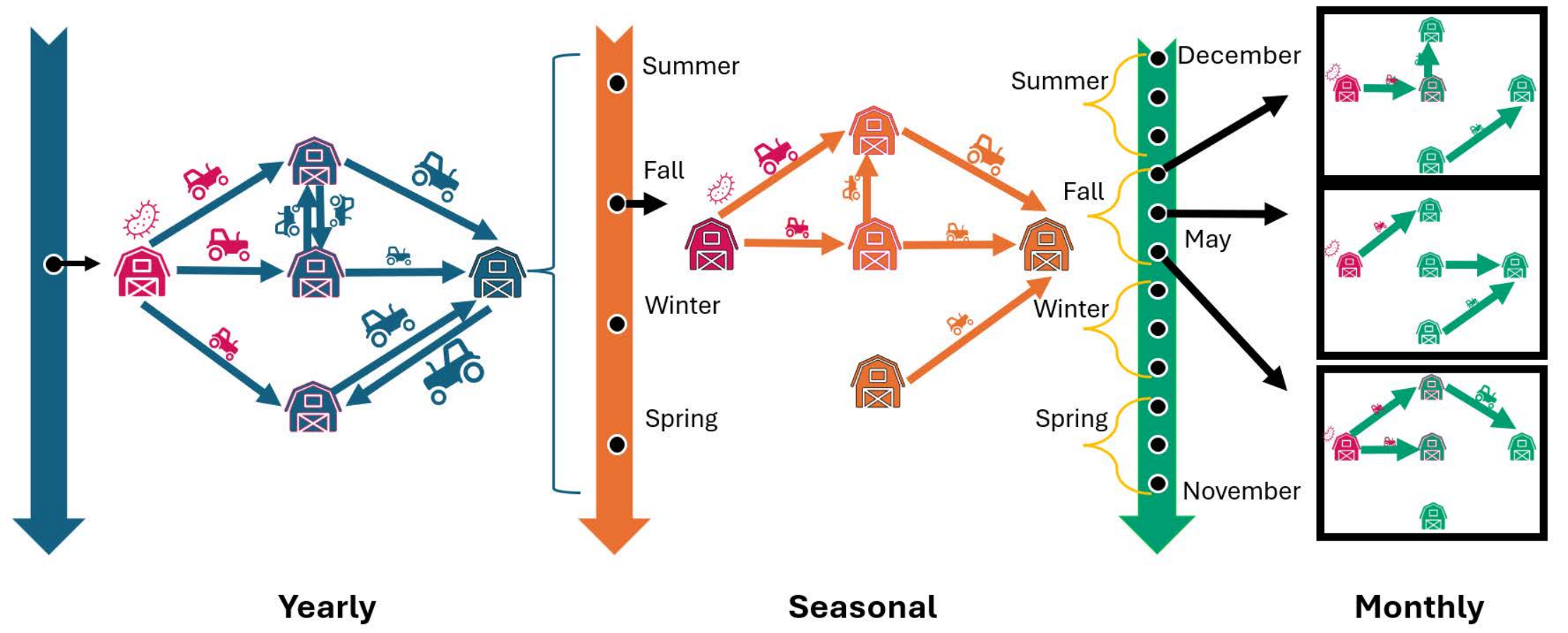}
    \caption{\textbf{Schematic representation of mobility networks.} Using the Onside movement dataset, we generate networks representing movements of workers between properties. In all networks, we measure the number of movements of people between properties, which vary based on the source and destination. A movement is directed in nature and is recorded when a person leaves a source property and checks into a destination property.  In this figure, the size of the tractor corresponds to the number of movements: the larger the tractor, the more movements occur. 
    Depending on the aggregation level of the data, we define either yearly, seasonal, or monthly networks.
    People moving between properties spread plant disease, and a higher frequency of movements between properties increases the probability of spreading. We note that starting the infection during different seasons or months can lead to different epidemic outcomes, since monthly movement networks differ. Infections are colored red showing how seeding infections during different seasons or months can lead to different epidemic pathways. }
    \label{fig:1}
\end{figure*}

%define network set 
  From the Onside data at our disposal, we extract all $P = 86,624$ movements recorded between January 1, 2022, and December 31, 2022 (see Figure~\ref{fig:1} for a schematic illustration). The data was pre-processed so time-stamps having a resolution of minutes are mapped to calendar months, since our epidemic model is calibrated on monthly timeseries data. As a consequence of such a rounding operation, the very same movement can appear multiple times in the data. Specifically, the $p$-th movement data is denoted by $z_p = (o_p,d_p,m_p)$, representing a person checking into property $d_p$ in month $m_p$, with $o_p$ representing the previous property where the very same person checked in. The indices $o_p $ and $d_p$ denote spatial units $1$ through $N$ ($N = 2,281$), while $m_p$ corresponds to one of the twelve calendar months (January–December).

We construct directed and weighted networks of movements for each month, excluding all spurious movements having the same source and destination properties. The generic element of the weight matrix for month $m$ is
\begin{equation}
\omega^{(m)}_{j \to i} = \frac{1}{30 \, \textrm{days} } \, \sum_{p = 1}^P  \delta_{o_p,j} \delta_{d_p,i}  \delta_{m_p, m} \; 
    \label{eq:weight_month}
\end{equation}
where the Kronecker delta function is such that $\delta_{x,y} = 1$ if $x = y$ and $\delta_{x,y} = 0$ otherwise. $\omega^{(m)}_{j \to i}$ quantifies the average number of daily movements between properties $j$ and $i$ observed in month $m$.

In a similar way, we can generate directed and weighted networks of movements for $4$ seasons: fall, winter, spring and summer. In such a case, the generic element of the weight matrix is
\begin{equation}
\omega^{(s)}_{j \to i} = \frac{1}{90 \, \textrm{days} } \, \sum_{p = 1}^P  \delta_{o_p,j} \delta_{d_p,i} \delta_{f(m_p), s} \; 
    \label{eq:weight_season}
\end{equation}
where $f(m)$ is the function that maps months to seasons, defined as follows: $f(m) = $ fall for $m = $ March, April and May,  $f(m) = $ winter for $m = $ June, July and August,  $f(m) = $ spring for $m = $ September, October, November, and $f(m) = $ summer for $m = $  December, January and February.  

The yearly version of the network is instead given 
\begin{equation}
\omega_{j \to i} = \frac{1}{365 \, \textrm{days} } \, \sum_{p = 1}^P  \delta_{o_p,j} \delta_{d_p,i}   \; 
    \label{eq:weight_year}
\end{equation}
where  all movements in the year are aggregated together.

% \cite{masuda2016guide,holme2019temporal}

A basic analysis of the constructed networks reveals that the volume of movements in the horticulture industry is highly seasonal; movements peak in the spring and summer seasons (highest in the months of February and November, shown in the inset of Figure \ref{fig:vary-c} C), and reach the lowest numbers in winter (June and July). This aligns with horticulture industry labor trends, where employment peaks in the summer and is smallest in the winter months~\cite{ffcove2025seasonal,timmins2009seasonal}.

% review how the table is created
 In Table ~\ref{tab:net_stats}, we show how the level of aggregation affects some basic network metrics. When calculating the metrics,  we only consider the properties having at least one non-zero-weight incoming or outgoing edge. We then calculate the average values over the set of all graphs generated according to a specific level of temporal aggregation for the number of nodes $\langle N \rangle$, the number of edges $ \langle E \rangle$, the first moment of the edges' weights $\langle \omega \rangle$, and the second moment of the edges' weights $\langle \omega^2 \rangle$.

 % go into some differences 
 By definition, the yearly network contains more nodes and edges than the monthly and seasonal networks. However, the weight of the connections is larger in the monthly and seasonal networks. This well aligns with an empirical understanding of how temporal networks exhibit ``bursty'' behavior with large amounts of activity occurring in a short period of time, followed by long periods of inactivity \cite{sheng2023constructing}.  
Estimates of the second moment highlight how, at all levels of aggregation, there is large heterogeneity in how the weights are distributed on the edges of the network.

% NETWORK STATISTICS

\begin{table}[!htb]
    \centering
    \renewcommand{\arraystretch}{1.2}
\setlength{\tabcolsep}{8pt}
    \begin{tabular}{|c|c|c|c|c|}
        \hline
        \textbf{Aggregation} & \textbf{$\langle N \rangle$} & {$\langle E \rangle$} & \textbf{$\langle \omega \rangle$} & \textbf{$\langle \omega^2 \rangle$}\\
        \hline
        Yearly & $2,281$ & $23,418$ & $0.01$ & $0.84$  \\ 
        \hline
        Seasonal  & $1,627$ & $8,124$   & $0.03$ & $0.80$   \\
        \hline
        Monthly  & $1,153$ & $3,413$   & $0.07$ & $0.86$    \\
        \hline
    \end{tabular}
     \caption{\textbf{Network statistics.} For each level of data aggregation, we report the corresponding average number of nodes $\langle N \rangle$, average number of edges $\langle E \rangle$, average value of the edges' weights $ \langle \omega \rangle$ (measured in number of daily movements), and the second moment of the edge weight  $\langle \omega^2 \rangle$.}
    \label{tab:net_stats}
\end{table}

\subsection{Metapopulation Model}

% defining populations for each each property

We are interested in developing a model for the spread of  Psa-V that is calibrated on a timeseries quantifying the monthly number of infected properties between February 2011 and March 2012, as collected by Greer and Saunders~\cite{greer2012costs}. To this end, we make the strong assumption that the $2022$ Onside data at our disposal can be used to generate networks of movements that are representative of both years $2011$ and $2012$. Specifically, we define the weight of the directed connection $j \to i$ at time $t$ as
\begin{equation}
w_{j \to i}(t) = \left\{
\begin{array}{ll}
\omega^{(m = g(t))}_{j \to i} & \textrm {monthly resolution}
\\
\omega^{(s = f(g(t)))}_{j \to i} & \textrm {seasonal resolution}
\\
\omega_{j \to i} & \textrm {yearly resolution}
\end{array}
\right.\; .
    \label{eq:weight_net}
\end{equation}
In the above expression, $g(t)$ is the function that maps the model's time $t$ to calendar months; $f(m)$ maps calendar months to seasons already used in Eq.~(\ref{eq:weight_season}).

The proposed metapopulation model combines two spreading mechanisms: within-property and between-property. Each property $i$, composed of $H_i$ hectares of land, is represented as a population of susceptible and infected hectares. We denote the fraction of infected hectares on the property $i$ at time $t$ as $I_i(t)$ and the fraction of susceptible hectares for property $i$ at time $t$ as $S_i(t)$. 
By definition, we have that
\[
S_i(t) + I_i(t) = 1, \forall i \in {1, \ldots, N} \textrm{ and } \forall t  \; .
\]

The level of transmission within a property is driven by the fraction of susceptible and infected hectares on that property (within-property transmission), and the importation of infection from other connected properties (between-property transmission). The equations describing the spread of the infection are:
% define mean-field equation
\begin{equation}
\begin{array}{ll}
\frac{d \, I_i(t)}{dt} = & \beta_{\mathrm{w}} \, I_i(t) S_{i}(t) \, \Theta \left[I_i(t) - I_i^{\min} \right] 
\\
& +\beta_{\mathrm{b}} \, S_{i}(t) \,  \sum_{j=1}^N w_{j \to i}(t)  \, I_{j}(t) \, \Theta \left[I_j(t) - I_j^{\min} \right] 
\end{array}
\; ,
\label{eq:cont}
\end{equation}
for $i = 1, \ldots, N$. $\Theta(x)$ is the Heavyside function defined such that $\Theta(x) = 1$ if $x > 0$ and 
$\Theta(x) = 0$ otherwise.
In Eq.~(\ref{eq:cont}), the force of infection within a property is denoted as $\beta_\mathrm{w}$, and the force of infection between properties is denoted as $\beta_\mathrm{b}$. 
%These are tunable parameters that we calibrate based on the available infection data (see Model Calibration). 
To be considered infected and contribute to the spreading dynamics, property $i$ must have a level of infection above the threshold $I_{i}^{\min}$ where $I_{i}^{\min} = \left(H_i \, 625 \, \mathrm{plants \, hect}^{-1}\right)^{-1}$; $625$ is a realistic estimate of the average number of plants per hectare~\cite{NZKGI2024}.
This threshold value is set to ensure that at least one plant in the property is infected before spreading occurs.

% review initial conditions for the equation 

Eq.~(\ref{eq:cont}) is a mean-field approximation of the ground-truth spreading dynamics~\cite{RevModPhys.87.925}. We assume that such an approximation well describes the average dynamical behavior of the system. We numerically integrate it using Newton's method (see Supplementary Materials for details). The initial condition at time $t = t_0$ is such that $I_i(t_0) = 0$ for all $i$ except for a single seed property $s_0$ for which $I_{s_0}(t_0) = 0.1$.As shown in the Supplementary Materials, the results of our analysis are not much affected by the initial level of infection imposed for the seed property.

% D-threshold info
Unlike livestock disease, it can be difficult to detect early-stage symptoms of plant disease, in part,  due to less routine monitoring. Thus, when monitoring the spread of the disease in our model, we require that  $I_i(t) > D$ for the property $i$ to be counted as infected at time $t$. We define the cumulative number of infections at time $t$ measured in our model as:
\begin{equation}
    O(t) = \sum_{i=1}^N \Theta \left[ I_{i}(t) - D \right] \; .
    \label{eq:infected}
\end{equation}

% describe transmission parameters in model
We stress that the specific choice for the value of the detection threshold $D$ does not affect how spreading unfolds, but only how spreading is monitored.
%We contrast $D$ to $I^{\min}_{i}$, which is the minimum percentage of hectares that must be infected for spreading to occur.  
$D$ should be interpreted as an effective detection threshold, which combines the time it takes to detect and report symptoms as well as biological incubation.

\subsection{Model Calibration}
 % OUTBREAK DATA
    \begin{figure}
        \centering 
        \includegraphics[width=1\linewidth]{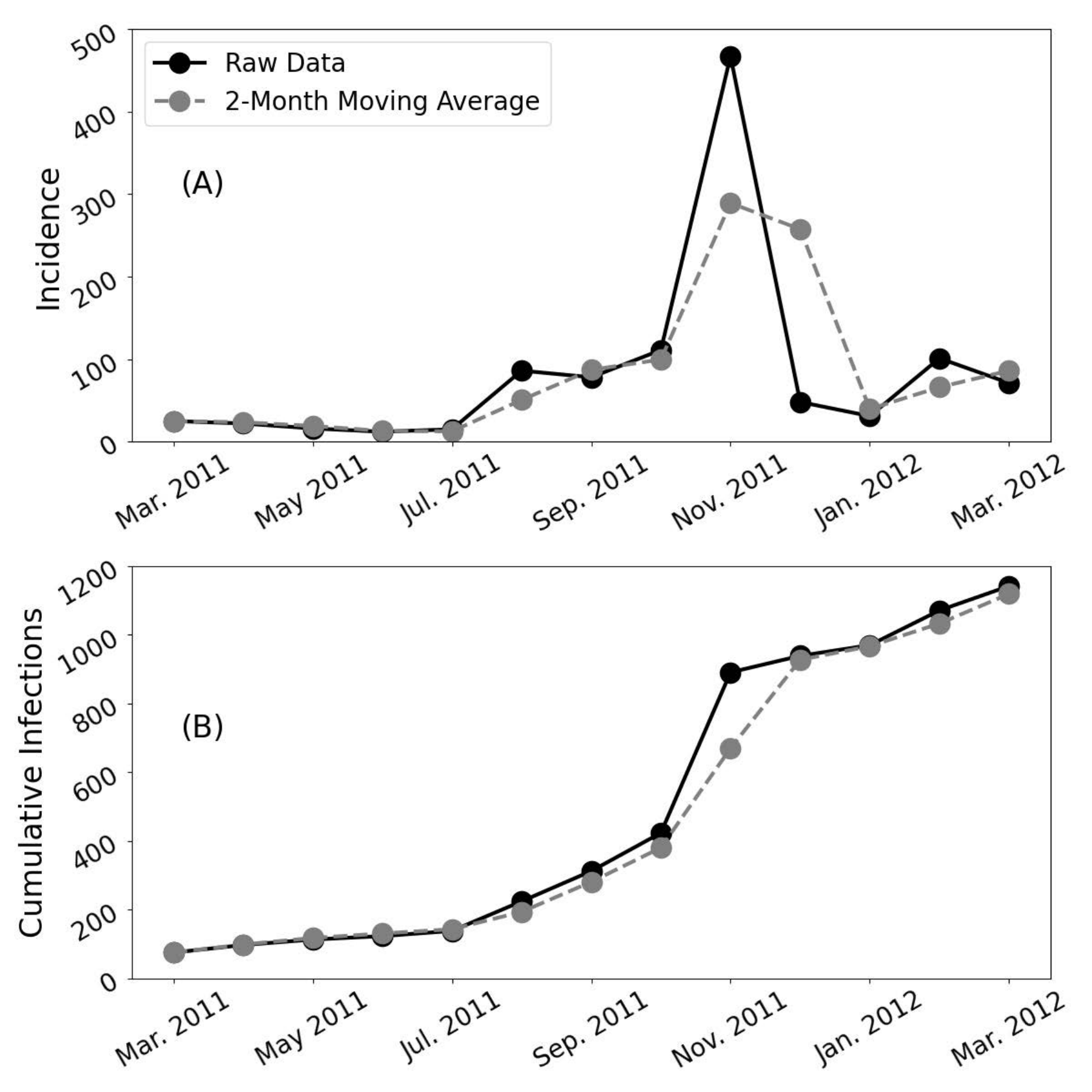}
        \caption{\textbf{Dynamics of the Psa-V epidemic in New Zealand.} 
        (A) Number of new monthly infected properties and (B) cumulative number of monthly infected properties. 
        The solid curve in both panels
        represents the raw data extracted from the report written by Greer and Saunders \cite{greer2012costs}. In this, the monthly infection data starts from February 2011 and ends in March 21st, 2012, but the actual outbreak started in November 2010 and has now become endemic in New Zealand. Thus, the leftmost point in the raw data
        corresponds to the number of properties that became infected between November 2010 and February 2011.
        The gray curve is a 2-month moving average of the raw timeseries. The moving average is calculated starting from March 2011. }
        \label{fig:real-data}
    \end{figure}

% review where the incidence data came

We use the data on the outbreak dynamics from a 2012 report by Greer and Saunders ~\cite{greer2012costs}. The report relies on infection records collected by Kiwifruit Vine Health (KVH), a biosecurity organization \cite{birnie2014lessons} responsible for protecting the kiwifruit industry from pests and disease. KVH  was established in December 2010 to lead the response to the Psa-V incursion. As the KVH disaggregated data is not publicly available, we extract the monthly values from the aggregated curve in the 2012 report using a plot digitizing tool~\cite{PlotDigitizer}. The number of infected properties per growing season was detailed in Rosanowski~\textit{et al.}~\cite{rosanowski2013quantification}; we use it to confirm the number of total infections in March 2012 in addition to the epidemic curve from Greer and Saunders~\cite{greer2012costs}. 

% review incidence curve 
The incidence curve shown in Fig.~\ref{fig:real-data} details the observed total monthly infections from 
February 2011 to March 2012. Since infections had occurred before February 2011~\cite{rosanowski2013quantification}, and the incidence in January 2011 and February 2011 was not available in the monthly dataset, 
the February data point is not included in our calibration procedure.

According to Rosanowski~\textit{et al.} \cite{rosanowski2013quantification}, the largest spikes in infection occurred in the spring during 2011/2012 growing season and the spring of the 2012/2013 growing season, with the summer season having the second largest number of Psa-V infected properties. Symptom emergence is greatest in the spring for Psa-V due to the warm humid conditions and increased rainfall \cite{froud2015review}. Although Psa-V symptoms are visually apparent once established, some orchards may have been infected earlier but only reported later due to biological latency, mild or ambiguous early symptoms, and limited awareness during the early outbreak phase. As an alternative reference curve for calibration, we also estimate a 2-month trailing moving average of the incidence curve. The moving average curve is calculated starting from February 2011 (see Supplementary Materials). 

% reviewing parameter choices in detail

The data used in our calibration procedure is given by $14$ monthly counts of newly infected properties. We denote this timeseries as $\vec{x} = (x_1, x_2, \ldots, x_{14})$, where we assume that data points are indexed in chronological order. This means that, for example, the index $c=1$ refers to February 2011 and $x_1$ is the value of the incidence curve in February 2011, whereas the index value $c = 14$ refers to March 2012 and $x_{14}$ is its corresponding value on the incidence curve. To compare the outcome of the metapopulation model with such real data, we discretize the timeseries of Eq.~(\ref{eq:infected}) by extracting a single value for each month from February 2011 to March 2012, i.e.,  $\vec{m} = (m_1, \ldots, m_{14})$. Each of these values corresponds to the cumulative number of infections measured at the end of the corresponding month. For example, $m_1$ is the cumulative number of infections measured in the model at midnight of March 1st,  2011; $m_{14}$ is instead the cumulative number of infections measured at midnight of April 1st, 2012. From the timeseries describing the cumulative number of infections, we immediately derive the incidence timeseries as $y_1 = m_1$ and $y_c = m_c - m_{c-1}$ for $c = 2, \ldots, 14$. 

The outcome of the metapopulation model depends on six tunable parameters, see Table \ref{tab:model_params}. In our calibration procedure, we assume that the dynamics starts on November 1st, 2010, thus effectively bringing down the number of free parameters to five~\cite{rosanowski2013quantification,greer2012costs}. More simply, our model runs for four months before we compare incidence to the real data.

For each level of temporal aggregation, we perform a grid search to find the best-fitting values for the other model parameters ($\beta_\mathrm{w}$,$\beta_\mathrm{b}$,$D$,$s_0$), which we detail in more depth in the Supplementary Materials. Potential configurations are filtered using the heuristic criterion detailed below in Equation \ref{eq:filter}. Note that we run separate calibrations for the different levels of temporal aggregation of the movement data. Also, note that, still relying on evidence from prior literature~\cite{rosanowski2013quantification,greer2012costs}, we allow seed nodes to be taken only from the set of properties in the Bay of Plenty, see Figure~\ref{fig:nz-seeds}, as the outbreak originated from that region. The goal of the calibration is to ensure that the resulting dynamics fits available incidence data with an acceptable level of uncertainty, a common goal of other disease models  \cite{merler2015spatiotemporal}.

% filtering criterion 
% PARAMETER DESCRIPTION TABLE
\begin{table}[t]
\centering
\caption{\textbf{Model Parameters and their description.} 
}
\begin{tabular}{ll}
\hline
\textbf{Parameter} & \textbf{Description}  \\
\hline
\hline
$R$ & Temporal resolution for the movement network \\
$\beta_{\mathrm{w}}$ & Within-property transmission rate  \\
$\beta_{\mathrm{b}}$ & Between-property transmission rate  \\
$s_0$ & Initial seed property \\
$t_0$ & Initial time of the spreading process \\
$D$ & Detection threshold   \\
\hline
\end{tabular}
\label{tab:model_params}
\end{table}
Given a configuration $\theta$ of the free model parameters and an aggregation $R$, we numerically integrate the Eqs.~(\ref{eq:cont}) of the metapopulation model to generate the incidence curve $\vec{y}(\theta) = (y_1(\theta), \ldots, y_{14}(\theta))$. Such a configuration is considered viable, i.e., $\theta \in V$, if  
\begin{equation}
 |x_c - y_c(\theta)| < \max \{ 30, 0.8 x_c \}
 \label{eq:filter}
\end{equation}
for all $c = 2, \ldots, 14$. The above is a heuristic criterion that allows us to select only configurations of the model's parameters whose outcome is compatible with the ground truth in the entire observation window for the data at our disposal. In the Supplementary Materials, we plot the percentage of viable configurations while varying these two heuristic values in a heatmap. From that, we chose $30$ and $0.8$  in the criterion above as they were the first two values with more than $0$ viable configurations for all aggregations, when using the raw data.  We use this criterion to prevent overfitting to any one specific month and filter any configurations that do not exhibit similar growth trends to the incidence data. Among all viable configurations, we identify the best configuration $\hat{\theta}$ as the one minimizing the root mean squared error (RMSE), i.e., 
 \begin{equation}
    \hat{\theta} =  \arg \min_{ \theta \in V } \sqrt{\frac{\sum_{c=2}^{14} (x_c - y_c(\theta))^2}{13}}
    \label{eq:opt}
 \end{equation}
 
Note that in both Eqs.~(\ref{eq:filter}) and~(\ref{eq:opt}), we start from the index $c=2$ thus excluding the first data point in the timeseries. This same procedure is also used to determine the values of the model parameters that best fit the 2-month moving average of the incidence curve, Table \ref{tab:best_params}. 

%The above procedure serves to estimate the best values of the parameters for the monthly, seasonal, and yearly versions of the spreading model. Those values are reported in Table~\ref{tab:best_params}. For each data type, we bold the configuration that minimizes RMSE best according to Equation \ref{eq:opt}.  

\subsection{Spatiotemporal characterization of the spreading dynamics}

% review distance metrics

Our metapopulation model of Eq.~(\ref{eq:cont}) allows us to keep track of the level of infection $I_i(t)$ of each property $i$ at time $t$. Similarly to what was described in the previous section, we discretize such a continuous timeseries into $14$ values $\vec{m} = (m_1^{(i)}, \ldots, m_{14}^{(i)})$ each value corresponding to the months between February 2011 and March 2012. We define the time to detect infection for property $i$, namely  $c^*_i$, as the minimum value of the index $c$ for which $m_c^{(i)} > D$. We finally measure the monthly detection distance $d_i^*$ between $i$ and the closest property $j$ such that $c^*_j = c^*_i - 1$. Since multiple properties can be detected at the same time $c^*$, we compute the minimum, mean, and maximum of the quantity $d^*$ for all properties that are first detected as infected at time $c^*$. 

Similarly, we estimate the distance between property $i$, first infected at time $c_i^*$, and each of the connected properties $j$ that are first infected at time $c_j^* = c_i^*-1$. We refer to this quantity as the contact distance $d_{i,j}$. For each value $c^*$, multiple values of the contact distance may be observed; thus, we compute the minimum, mean, and maximum of the contact distance for all properties that are first detected as infected as functions of $c^*$. 

These distance metrics allow us to track whether the outbreak distance radically changes over time. The contact distance metric tells us the distance that newly infected properties have from their infectors, while the outbreak distance metric sheds light on the distance that newly infected properties have from one another. The latter metric gives information on how the front of the epidemic expands, month by month.

\section{Results}

\subsection{Replicating observed spreading dynamics}

\begin{table}[ht]
\centering
\renewcommand{\arraystretch}{1.2}
\setlength{\tabcolsep}{8pt}
\begin{tabular}{|l|l|c|c|c|}
\hline
\textbf{Data} & \textbf{Aggregation} & $\boldsymbol{\beta_b}$ & $\boldsymbol{\beta_w}$ & $\boldsymbol{D}$ \\
\hline
\multirow{3}{*}{Raw}
 & Monthly  & 0.03 & 0.09 & 0.4 \\
 & \textbf{Seasonal} & 0.03 & 0.08 & 0.8 \\
 & Yearly   & 0.04 & 0.07 & 0.8 \\
\hline
\multirow{3}{*}{Moving Average}
 & \textbf{Monthly}  & 0.02 & 0.10 & 0.4 \\
 & Seasonal & 0.03 & 0.07 & 0.8 \\
 & Yearly   & 0.02 & 0.10 & 0.4 \\
\hline
\end{tabular}
\caption{\textbf{Best-fitting parameter values.} Best-fitting parameter values for each temporal aggregation using the raw and moving average incidence data. 
We use bold-face fonts to highlight the best-fitting aggregation level for each type of reference curve.
}
\label{tab:best_params}
\end{table}

%Table \ref{tab:best_params} presents the parameter estimates for the model, categorized by data aggregation level and the type of reference timeseries (raw incidence vs. moving average). Two key observations emerge from these results. First, the estimates remain highly robust across different aggregation levels, with the within-property transmission rate consistently approximately three times greater than the between-property rate. Second, as illustrated in the Supplementary Materials, the posterior distributions are well-defined and unimodal, centered sharply around their respective modes.

In Table \ref{tab:best_params}, we report the values of the best estimates of our model. They are grouped based on the level of data aggregation; also, we display separate results depending on whether the reference timeseries is based on the raw incidence data or their moving average. Two nice aspects are apparent from our estimates. First, their values are very robust across the various levels of aggregation, with the within-property contribution being about three times larger than its between-property counterpart. Second, as show in the Supplementary Materials, their posterior distributions are well peaked around their mode.

\begin{figure}[h]
    \centering
\includegraphics[width=\linewidth]{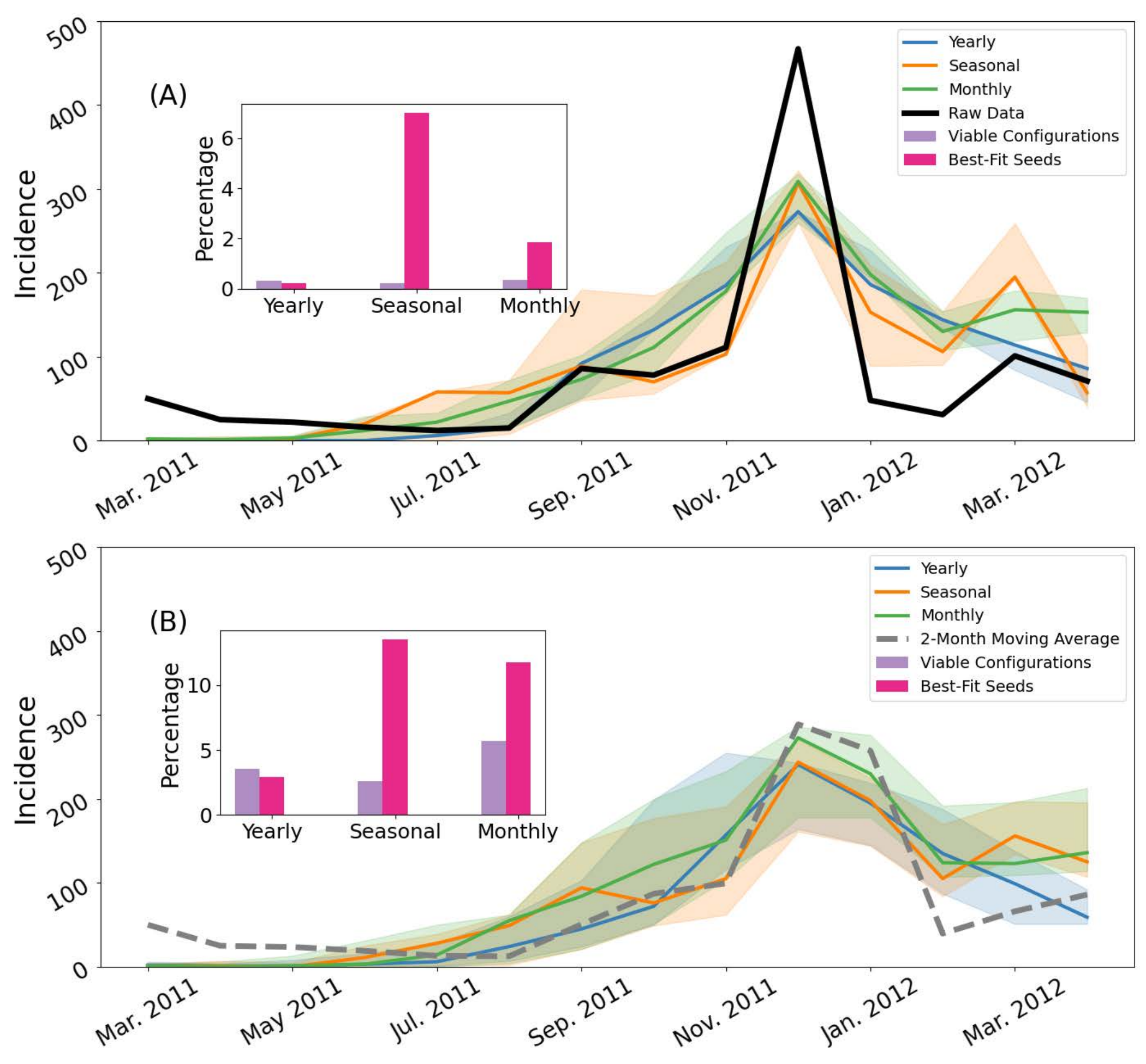}
    \caption{\textbf{Best-fitting epidemic models.} 
    (A) We plot the incidence timeseries obtained from the epidemic model when setting its parameter values to their corresponding best estimates. Different colors refer to different levels of temporal aggregation of the movements between properties: yearly (blue), seasonal (orange), and monthly (green). The shaded areas define the uncertainty associated with each of these curves, computed by taking the minimum and maximum values of the incidence measured over the viable seeds for the best-fitting values for $\beta_b,\beta_w,D$ . Model's curves are compared against raw data (black). The inset displays the percentage of viable configurations for the parameters in violet, as well as the percentage of unique seeds appearing in the best-fitting configuration when holding $\beta_b,\beta_w,D$ to their best fitting value. This analysis was conducted for the various levels of temporal aggregation that can be in the model.
    (B) Same as in (A), but obtained when fitting the model against the 2-month moving average incidence curve (gray dashed).  }
    \label{fig:best-fits}
\end{figure}

% overview of best fit figur4e  
 We visually compare the incidence curves from the raw data and the model estimates corresponding to the best fit for the three different levels of temporal aggregation in Figure \ref{fig:best-fits}A. The bands for each of the model's curves represent the uncertainty of the incidence estimates due to the choice of the seed $s_0$ when using the best-fitting parameter values for $\beta_w,\beta_b,D$ given the temporal aggregation and incidence curve. Within the insets of Figure \ref{fig:best-fits}, we compute the total percentage of viable configurations in $V$ and the percentage of unique seeds observed given the best-fitting values for $\beta_b,\beta_w$ and $D$ for all levels of temporal aggregation.  A similar analysis is considered in Figure \ref{fig:best-fits}B, the model is fit against the 2-month moving average of the incidence curve. 

% need for temporal network 

The results shown in Figure \ref{fig:best-fits} illustrate that the model is able to capture epidemic dynamics well for all levels of temporal aggregation. The model seems to better fit the 2-month moving average of the incidence curve rather than its raw version. The difficulty of fitting well the raw incidence timeseries arises from the extreme steepness of the outbreak peak, with doubling time being less than the available data time step (1 month). This highlights the need for higher temporal disaggregation of infection data for the calibration of the proposed model when highly transmissible pathogens are analyzed.

% state best network type

Of all levels of temporal aggregations, the seasonal configuration provides the best fit, i.e., smallest RMSE [Eq.~(\ref{eq:opt})], for the dynamics of the observed incidence (Fig. \ref{fig:best-fits}A). It also has the largest percentage of seeds for its best-fitting configuration, but the smallest percentage of viable configurations. For the 2-month moving average epidemic curve (Fig. \ref{fig:best-fits}B), the monthly network provides the best fit. Compared to the observed incidence, the percentage of viable configurations and best-fitting seeds is greater than for the 2-month moving average across temporal aggregations. The root mean-squared error values are smaller for the best-fitting configurations for the moving average compared to the raw data. This indicates that the model struggles less with smoother dynamics when doubling time is comparable (or preferably) shorter than the temporal observational scale. 

% summarize main result of figure 

The model is well calibrated in terms of outbreak timing, with the correct peak month being captured perfectly by all levels of temporal aggregations, both for the raw incidence curve and its moving average. Additionally, the best-fitting seasonal model calibrated for raw incidence captures not only the peak that occurs in November 2011 but also the second incidence increase that occurs in February 2012.

% figure for best fitting curves

\subsection{Spatial Spread}

% overview of distance type figures 

\begin{figure}
    \centering
    \includegraphics[width=\linewidth]{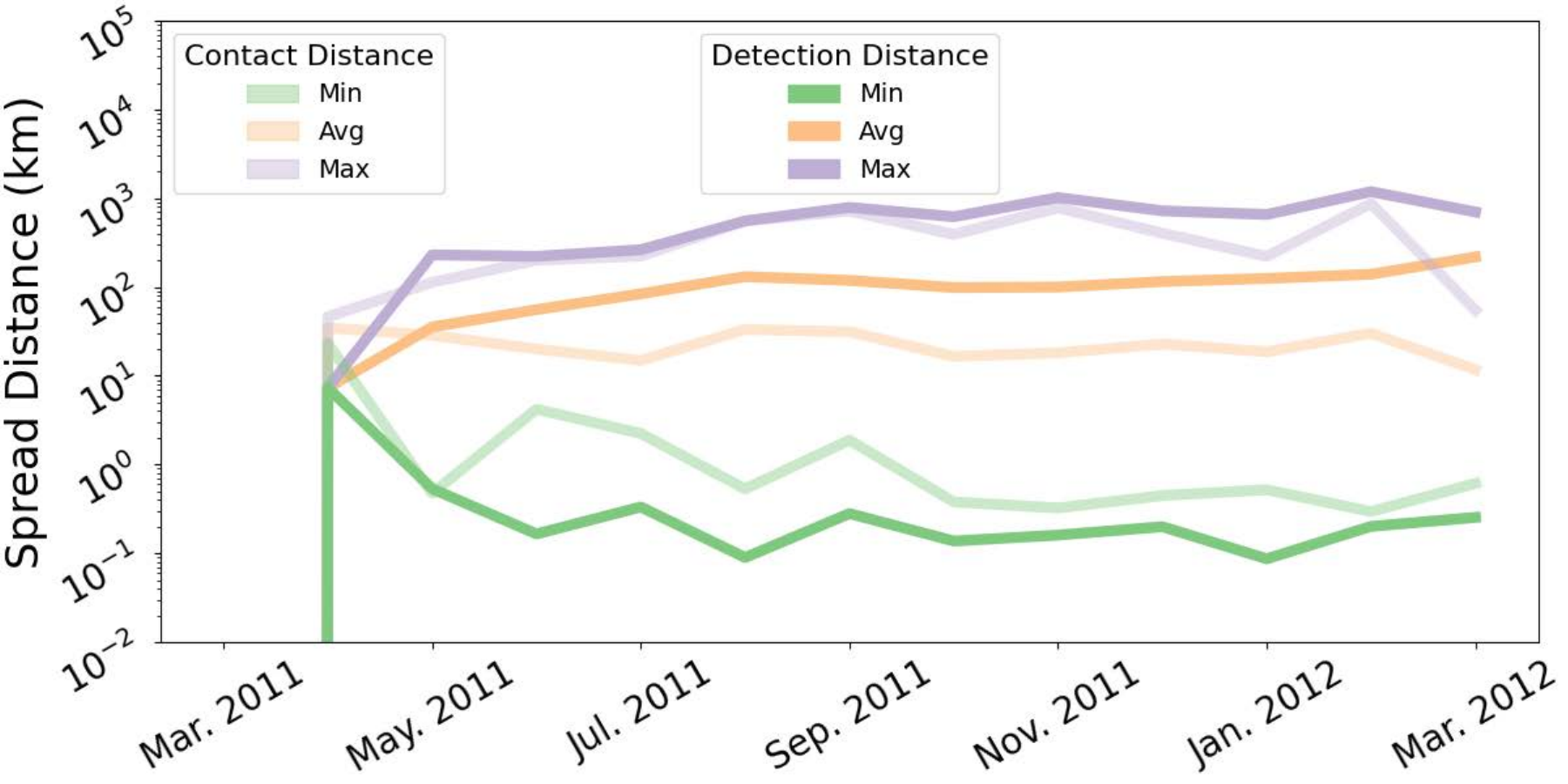}
    \caption{\textbf{Spatio-temporal characterization of the spreading.} 
    Two distance metrics are compared using three different statistics: the mean, minimum, and maximum. The detection-distance metric is calculated by comparing the distances of all properties that are infected within the same month. The contact distance is calculated by finding the average distance of a newly infected property’s contacts. A contact is defined as a previously infected property that has an outgoing link to the newly infected property. We define a newly infected property as one that is detected as infected, not simply if it can transmit infection. 
    }
    \label{fig:dist-types}
\end{figure}

As seen in Figure \ref{fig:dist-types}, most spreading events are initially local. Over 90\% of all newly infected properties are less than 10 km from their closest infectors (see SM), illustrating the dominance of local spread. However, rare long-distance events pull the average distance between newly infected properties and another closest infected property into the range of 10–100 km (Fig. \ref{fig:dist-types}). This feature is a direct result of the movement network's spatial structure, with most movements between properties ($71.3\%$) being short-range ( $<10$km), $12.2\%$ of movements being $> 20$km and rare cases reaching $1,000$ km (the CCDF of the movement distribution is shown in the SM). Moreover, the importance of such movements for transmission increases as the outbreak continues. This is illustrated by the average distance from the closest infected property consistently increasing each month. The heterogeneity of the distance between infected contacts and newly infected nodes also increases with time due to two factors: (i) the number of close susceptible nodes decreases, and (ii) the pathogen reaches close but less connected properties (a smaller number of movements). This leads to initially small and local outbreak spreading far beyond its origin. In the SM, we plot how the cumulative infection distance from the initial seed grows over time. Tracking the outbreak from the source of infection, the infection spreads more than $1,000$ km from where it is seeded. We observe that growth from the seed occurs in a staggered manner, where every $2-3$ months there is an increase in distance from the seed property.

 % FIGURE: Distance Types 

\subsection{Experimental Analysis}

% review what we did in sensitivity analysis

\subsubsection{Detection threshold}
% FIGURE - VARYING D THREHSOLD
\begin{figure}
    \centering
    \includegraphics[width=\linewidth]{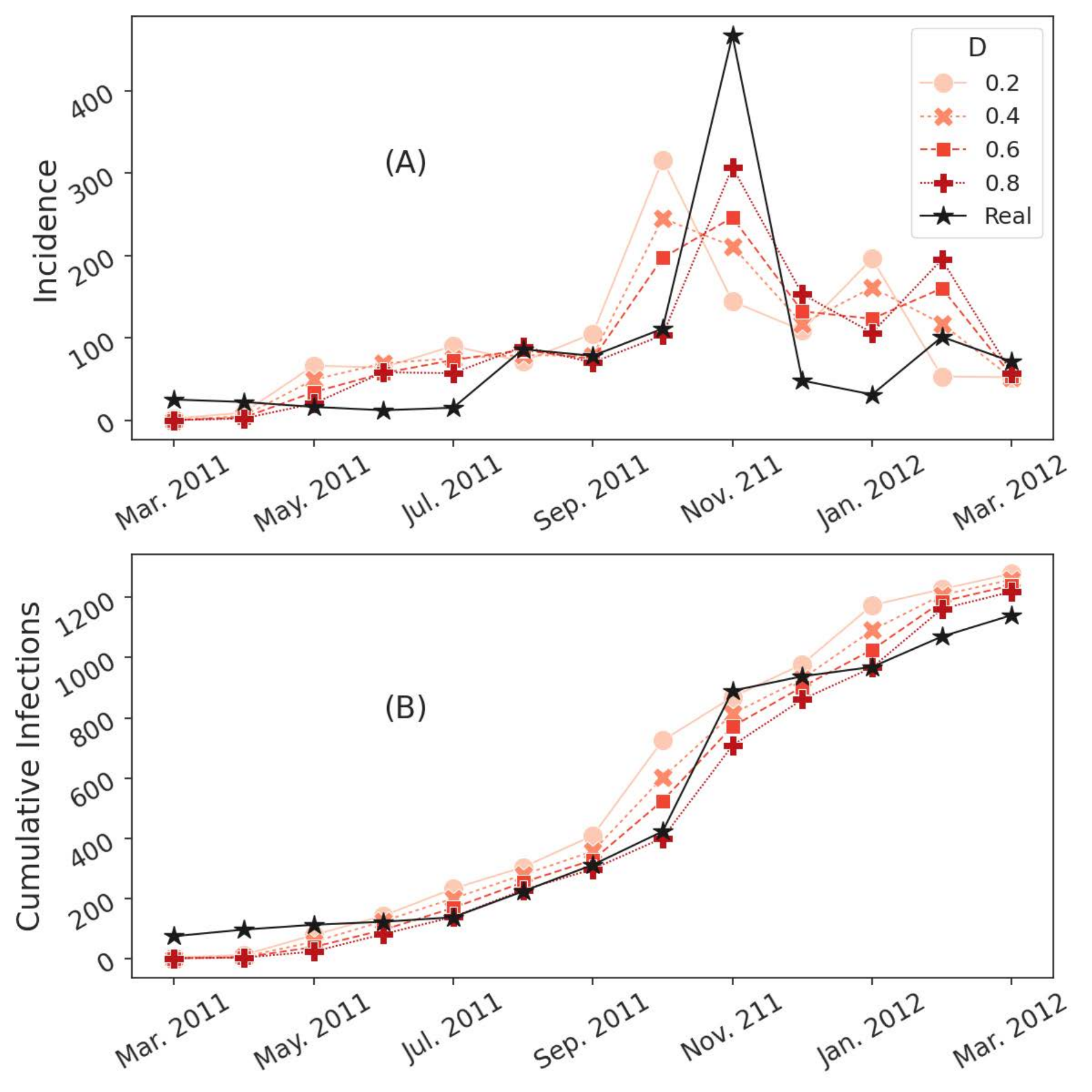}
    \caption{ 
    \textbf{Dependence of the model's outcome on the detection threshold.}
    We vary the detection threshold $D$ and plot the effect this has on the incidence curve in panel (A) and on the cumulative total infections shown in panel (B). The best fitting parameter configuration i’s $D = 0.8$ for the seasonal network.
    In panel (A), we plot the raw incidence curve as a reference, and in panel (B), the raw cumulative infection data are plotted. For all graphs, the seasonal network is used as well as the corresponding parameters of the best fit, Table~\ref{tab:best_params}.}
    \label{fig:vary-d}
\end{figure}

% findings of detection threshold result 

The detection threshold $D$ is the minimum fraction of infected hectares that must be present on a property for it to be considered infected. The larger $D$ is, the more hectares must become infected for a property to be observed as infected, see Eq.~(\ref{eq:infected}). Figure \ref{fig:vary-d} shows the temporal dynamics of the outbreak for different $D$ values; we use the network constructed using seasonal aggregation of movement data, and set all the model's parameters to their best-fitting values (Table~\ref{tab:best_params}, and Figure \ref{fig:best-fits}A). Decreasing $D$ effectively allows us to observe outbreak occurrence earlier.  

% analysis of D figure

Despite the peak incidence shifting by one month, the model illustrates similar temporal dynamics for different detection thresholds. The magnitude of incidence is similar when comparing the smallest and largest values, and the final outbreak size only differs by $2.7\%$, decreasing from $56.1\%$ of properties for $D = 0.2$ to $53.4\%$ for $D = 0.8$. This illustrates that, without implemented interventions, early detection does not substantially affect the observed final size of the outbreak or the height of the peak incidence.

\subsubsection{Importation month}

% review importation month figure 

All the above analysis was conducted using the realistic assumption that the importation month was November 2010. Now, we consider hypothetical scenarios where the time of pathogen importation may occur at the beginning of each calendar month during the year. For each month, we integrate the equations of the metapopulation model using the best estimates of the model's parameters for the monthly aggregation, see Table~\ref{tab:best_params}. To make the various scenarios comparable, the equations are integrated for the same total amount of time of $14$ months. When the hypothetical importation month is set to November, we obtain very similar initial conditions to those used in all previous results, though we only run this experiment for 365 days compared to the calibrated model. 

\begin{figure*}[!htb]
    \centering
    \includegraphics[width=\linewidth]{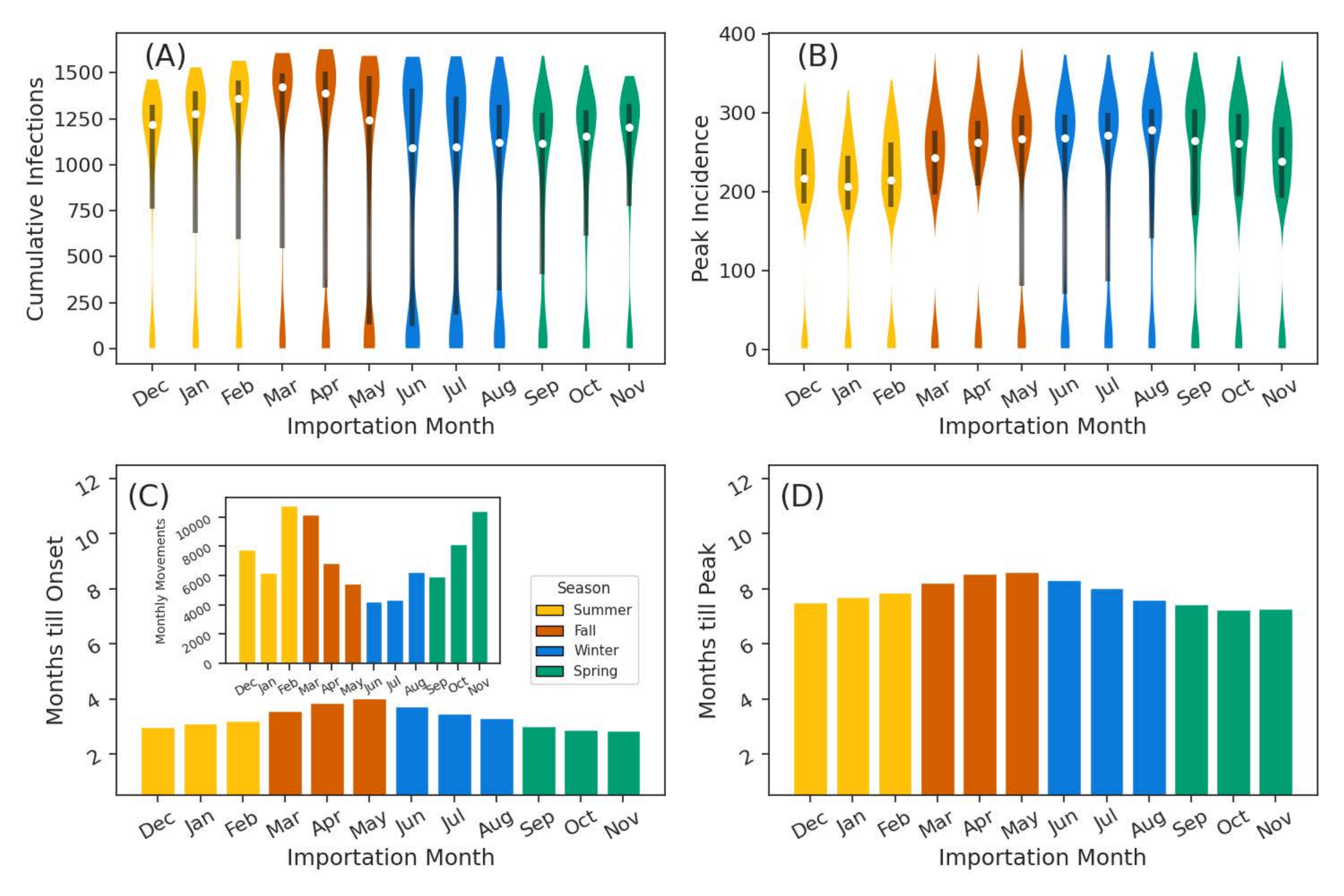}
    \caption{
    %\textbf{Epidemic start month affects the dynamics of the outbreak} 
    \textbf{Dependence of the model's outcome on importation month.}
    (A) The distribution of the final outbreak size (violin plots) for each importation month. Black boxes show the 50\% inter-quartile range. White circles show the median value. 
    %The distribution for each month is generated from all seeds used for calibration. 
    (B) As in (A), but for the distribution of peak incidence. (C) The mean number of months from importation until the month of peak incidence. In the inset, the total number of movements per month is plotted. (D) The mean number of months from importation until the onset (time when 2 infected farms are detected). Colors of bars and violins denote seasons.} 
    \label{fig:vary-c}
\end{figure*}

To evaluate the impact of the timing of pathogen importation, we consider four different metrics characterizing the spread of the epidemic: (i) the overall number of infections (Fig.~\ref{fig:vary-c}A), (ii) the peak incidence value (Fig. ~\ref{fig:vary-c}B) as well as (iii) the time when such a peak is reached (Fig.~\ref{fig:vary-c}D), (iv) the effective start of the epidemic which we define as the first instant of time when at least two properties are detected as infected (Fig.~\ref{fig:vary-c}C). The number of total movements per month is shown as well (inset of Fig.~\ref{fig:vary-c} C), and we compare the cumulative infections/incidence trends against the number of movements per month. 
All metrics were calculated by seeding infection from properties within the Bay of Plenty. The variation in distributions occurs due to the differences in the first movement network associated with the different importation months.   

% Panel (A)
Every month, there is a fraction of seeds where the final outbreak size is smaller than $250$ infections. In May-July, this fraction increases together with an overall decrease in outbreak size heterogeneity, resulting in a smaller median final outbreak size. Our results show that the biggest outbreak sizes are observed when importation occurs in March and April, i.e., the months with the maximum number of movements.

% alternative description of peak incidence  - Maria
We observe that importations in the fall and winter months lead to a higher incidence at the peak of the outbreak, with the highest variability in May through August - similar to the cumulative infections distributions. There exists a consistently higher probability of the outbreak either not taking off or remaining very small in winter, reflected in similar trends for the final outbreak size and peak incidence. 
However, for the combination of epidemiologic parameters and importation seeds resulting in bigger outbreaks, peak incidence tends to be highest for winter importations and lowest for summer importations. 
With the median final outbreak size being the highest for fall importations and lowest for winter importations, this suggests different spatiotemporal dynamics for the importations in different seasons.

% go over timing trends, Panels (C) and (D)

Finally, we see that the onset ($2-4$ months) and peak ($7-8$ months) times are almost constant irrespective of the value of the importation month.  
We also find that it takes longer for outbreaks to peak (Fig. \ref{fig:vary-c}B) and start (Fig \ref{fig:vary-c}C) if they begin in the fall or winter seasons.

\section{Discussion}

% overview of research done
We use a metapopulation model that incorporates human movements to simulate plant disease spread.  Our model reproduces the dynamics of the Psa-V outbreak that occurred starting from 2010 in New Zealand,  demonstrating that the movements of people can be effectively utilized to model the spread of plant disease. The metapopulation model relies on underlying mobility networks generated by aggregating movement data over time windows of variable length, i.e., monthly, seasonal, and yearly time windows. While all of them approximate the infection dynamics well, the seasonal network fits the infection dynamics from the raw data best. This occurs since farm work is seasonal, so a seasonal resolution is sufficient to capture most movement complexity. 

Results from our model provide critical quantitative insights for biosecurity. We show that disease mostly spreads locally between properties, within a radius of about $10$ km, a finding supported by the literature \cite{rosanowski2013quantification}.
However, nationwide outbreaks can quickly occur due to rare long-distance movements leading to sudden jumps in spread distance. 
Due to this combination of short-range spreading events and rare long-distance jumps, plant disease outbreaks always have the potential to spread over large distances if the disease is not detected quickly enough.

% discuss experimetnal analysis results 
Our model also allows us to test the sensitivity of disease spreading to initial conditions. Detecting infection on a property depends on the intensity of monitoring efforts, and we find that increasing sensitivity to detecting infection could represent an effective surveillance strategy. Also, we find that the median size of the outbreak increases if disease is imported during months with many movements. At parity of importation time, the specific location where the disease is imported contributes to the large variability in epidemic severity. Some seed properties generate only small and local outbreaks, and some others cause system-wide epidemic events.

% detail novelty of using human movements

This work is one of the first to use the movements of people to simulate the spatio-temporal spread of plant diseases. Other works have focused on how environmental factors \cite{brown2002aerial,cendoya_individual-based_2024} or farm proximity \cite{strona2017network} affect plant disease spread. By contrast, we show that the movement network of people strongly affects the size and speed of an outbreak, with epidemic dynamics changing depending on the initial seed location. Here, we directly harness the movement of people to model disease spread, allowing surveillance strategies to be directly related to real movement patterns. If a novel plant disease begins to spread in New Zealand, the calibrated model could be used as a helpful tool to test different potential scenarios. 

% detail importance of modeling temporality, impact on surveillance

Worker movements are tied to the growing season, so the density of movements per property changes based on how busy properties are by season. This highlights the importance of modeling how the movement network changes in time, since the temporality strongly affects the distribution of movements. The horticulture industry could leverage this knowledge to use different outbreak management strategies for different seasons. More generally, this model could help forecast a plant disease outbreak in the short-term, since the calibration grounds the estimates in a past real-world outbreak, help decision makers plan for new outbreaks, or assess options during an ongoing one. 

% talk about limitations

While our model can provide insights into the impact of human movements on the spread of plant disease, our work has several limitations. Our model has not been extended to account for the impact of specific climatic variables such as humidity, rainfall, and wind. For parameter identifiability purposes, we focus our work on human mobility networks; however the model is flexible enough to adapt a more complex mechanism for within and between property spread.

Additionally, the monthly observed infection data that we used to calibrate the model were affected by active interventions that were occurring at the time to prevent Psa-V spread \cite{greer2012costs,birnie2014lessons}. The infection data was collected based on symptom emergence, rather than the identification of infection.
Further access to the individual property level data was limited, therefore model validation was based on aggregated monthly infection data. 

In a similar fashion, access to daily or weekly incidence data was limited. Such access would allow for aggregation of the movement data at finer temporal resolution. 

When calibrating the model, we utilized the full New Zealand horticulture movement network to simulate the spread of Psa-V, a kiwifruit disease. The 2010 outbreak was larger than the number of kiwifruit growing properties in the horticulture movement network. Utilizing only present day kiwifruit properties would result in underestimation of the historical outbreak data. The full horticulture movement network was used to give a fuller representation of the mobility network compared to only using the kiwifruit network.

When evaluating the calibrated model, we observe that our model underestimates the full magnitude of the peak incidence for the raw data across all aggregations. In contrast, our model well captures the peak incidence for the moving average data across all aggregations. This suggests that model performance is more sensitive to data type than aggregation choice. 

While this study effectively illustrates the calibration pipeline using a straightforward filtering and RMSE  approach, the framework is designed to be adaptable, allowing future research to incorporate traditional inference methods like Markov-chain Monte-Carlo methods \cite{chowell2017fitting} or particle filtering\cite{yang2014comparison}.

% now, future work 
This model offers several promising avenues for future research. A primary focus is identifying which properties function most effectively as sentinel nodes \cite{bajardi2012optimizing,colman2019efficient} to ensure the timely detection of outbreaks. Beyond surveillance, this modeling framework can also be used to evaluate various containment strategies to mitigate disease spread once an outbreak is identified. 

\section{Data Availability}
All the code and data needed to reproduce the results of the study are available on Github here: \url{https://github.com/vakrao/net_plant_spread}. Due to the potential identifiability of individual properties, the latitude and longitude coordinates associated with each property will not be released publicly. Access to the location data should be requested from Onside.    
\section{Acknowledgments}
This project was supported by the Onside company via a collaborative grant. We would like to thank Reza Shoorangiz and Zach Hitchcock from Onside for their help. The company had no role in study design, data analysis, the decision to publish, or any opinions, findings, and conclusions or recommendations expressed in the manuscript. F.R. acknowledges partial support by the Air Force Office of Scientific Research under award number FA9550-24-1-0039.
\bibliography{refs}

\end{document}

\title{Modeling plant disease spread via high-resolution human mobility networks \\
Supplementary Materials}

\author{Varun K. Rao}
\affiliation{Center for Complex Networks and Systems Research, Luddy School of Informatics, Computing, and Engineering, Indiana University, Bloomington, Indiana 47408, USA}
\affiliation{School of Public Health, Indiana University, Bloomington, Indiana 47408, USA}
\author{Ryan Higgs}
\affiliation{Onside, Christchurch,  New Zealand.}
\author{Hautahi Kingi}
\affiliation{Onside, Christchurch,  New Zealand.}
\author{Filippo Radicchi}
\affiliation{Center for Complex Networks and Systems Research, Luddy School of Informatics, Computing, and Engineering, Indiana University, Bloomington, Indiana 47408, USA}
\author{Santo Fortunato}
\affiliation{Center for Complex Networks and Systems Research, Luddy School of Informatics, Computing, and Engineering, Indiana University, Bloomington, Indiana 47408, USA}
\author{Maria Litvinova}
\email{malitv@iu.edu}
\affiliation{School of Public Health, Indiana University, Bloomington, Indiana 47408, USA}
%TC:incbib

\maketitle

\section{Numerical Integration of the Mean-Field Approximation}

Denote with $I_i(t)$ the fraction of hectares that are infected in farm $i$ at time $t$, and with 
$S_i(t)$ the fraction of hectares that are not infected in farm $i$ at time $t$. We clearly have
$I_i(t) + S_i(t) = 1$ for all $i= 1, \ldots, N$ and all $t \geq 0$.

The farm-based mean-field equations for the SI dynamics are
\begin{equation}
\begin{array}{ll}
\frac{d \, I_i(t)}{dt} = & \beta_{\mathrm{w}} \, I_i(t) S_{i}(t) \, \Theta \left[I_i(t) - I_i^{\min} \right] +
\beta_{\mathrm{b}} \, S_{i}(t) \,  \sum_{j=1}^N w_{j \to i}(t)  \, I_{j}(t) \, \Theta \left[I_j(t) - I_j^{\min} \right] 
\end{array}
\; ,
%\frac{d \, I_i(t)}{dt} = \beta_{w} \, I_i(t) S_{i}(t) + \beta_{b} \, S_{i}(t) \,  \sum_{k=1}^N w_{k \to i}(t,Z)  I_{k}(t) \; 
\label{eq:cont}
\end{equation}
where $\beta_{w}$ and $\beta_{b}$ are the transmission rates within and between farms, respectively.  $\Theta(x)$ is the Heavyside function defined such that $\Theta(x) = 1$ if $x > 0$ and 
$\Theta(x) = 0$ otherwise. To be considered infected and contribute to the spreading dynamics, property $i$ should have a level of infection that is above the threshold  $I_{i}^{\min} = \left(H_i \, 625 \, \mathrm{plants \, hect}^{-1}\right)^{-1}$, where $625$ is a realistic estimate of the average number of plants per hectare~\cite{NZKGI2024}. $\omega^{(m)}_{j \to i}$ quantifies the average number of daily movements between properties $j$ and $i$ observed in month $m$.
%$\beta$s  are the "free" parameters that should be calibrated based on the available data (from literature or an ongoing outbreak) and will absorb the differences in the absolute values between $I_i(t)$ and $\sum_{k=1}^N w_{k \to i}  I_{k}(t)$.
%$\beta_{b}$ should always be small enough so that $( \beta_{w} \, I_i(t) + \beta_{b} \,  \sum_{k=1}^N w_{k \to i}  I_{k}(t) ) < 1 $ for all $i$.

We can also write the above as a finite-difference equation
\begin{equation}
I_{i}(t+dt) = I_{i}(t) + \beta_{w} \, dt \, \, I_i(t)  S_{i}(t)\Theta \left[I_i(t) - I_i^{\min} \right] + \beta_{b} \, dt \, \, S_{i}(t) \,  \sum_{j=1}^N w_{j \to i}(t)  \, I_{j}(t) \, \Theta \left[I_j(t) - I_j^{\min} \right] 
\label{eq:fd}
\end{equation}
%Now, 
%we numerically integrate
To perform the numerical integration, we require that  

\begin{equation}
    \beta_{w} \, dt \, \, I_i(t)  S_{i}(t)\Theta \left[I_i(t) - I_i^{\min} \right]  +  \beta_{b} \, dt \, \, S_{i}(t) \,  \sum_{j=1}^N w_{j \to i}(t)  \, I_{j}(t) \, \Theta \left[I_j(t) - I_j^{\min} \right]  \leq 1 - I_i(t)\; .
\end{equation}

%The condition is on the time increment $dt$ that can not be arbitrarily large, rather
leading to the constraint for the time increment:
\begin{equation}
    dt  \leq \frac{1 - I_{i}(t)}{\beta_{w} I_i(t)  S_{i}(t) \Theta \left[I_i(t) - I_i^{\min} \right] + \beta_{b}  \, \, S_{i}(t) \,   \,  \sum_{j=1}^N w_{j \to i}(t)  \, I_{j}(t) \, \Theta \left[I_j(t) - I_j^{\min} \right]} \; .
\end{equation}
Such a condition means that Eq.~(\ref{eq:fd}) can effectively be used to numerically integrate Eq.~(\ref{eq:cont}).
Since the above must be true for all farms, we need to impose

\begin{equation}
     dt  \leq \min \, \bigg( T_{\mathrm{inc}} ,\min_{i=1, \ldots, N} \, \frac{1 - I_{i}(t)}{\beta_{w} I_i(t)  S_{i}(t) + \beta_{b}  \, \, S_{i}(t) \,  \sum_{k=1}^N \frac{w_{k \to i}}{\tau}  I_{k}(t)} \; \bigg).
\end{equation}
     
where $T_{\mathrm{inc}} = 1$ to generate at least one data point per day.
%is the minimum daily time increment, so we set $T_{inc} = 1.$

\section{Infection Data}
The number of monthly cumulative infections and incidence from the 2010 Psa-V outbreak in New Zealand was digitized from Greer and Saunders \cite{greer2012costs} and is listed in Table \ref{tab:incidence_ma_simplified}. Raw data values are shown as well as the values for the trailing 2-month moving average. Using this incidence data we calibrated our model.
\begin{table}[H]
\centering
\caption{Monthly raw counts and two-month moving average (MA(2)) incidence and cumulative infections.}
\label{tab:incidence_ma_simplified}
\renewcommand{\arraystretch}{1.2}
\setlength{\tabcolsep}{6pt}
\begin{tabular}{|c|cc|cc|}
\hline
\textbf{Month} 
& \multicolumn{2}{c|}{\textbf{Raw Counts}} 
& \multicolumn{2}{c|}{\textbf{2 Month Moving Average}} \\
\cline{2-5}
 & Incidence & Cumulative 
 & Incidence & Cumulative \\
\hline
Feb.\ 2011 & 50  & 50   & 50  & 50   \\
Mar.\ 2011 & 25  & 75   & 25  & 75   \\
Apr.\ 2011 & 22  & 97   & 23.5  & 98.5  \\
May.\ 2011 & 16  & 113  & 19  & 117.5  \\
Jun.\ 2011 & 10  & 123  & 13  & 130.5  \\
Jul.\ 2011 & 15  & 138  & 12.5  & 143  \\
Aug.\ 2011 & 86  & 224  & 50.5  & 193.5  \\
Sep.\ 2011 & 88  & 312  & 87  & 280.5  \\
Oct.\ 2011 & 111 & 423  & 99.5 & 380  \\
Nov.\ 2011 & 467 & 890  & 289 & 669  \\
Dec.\ 2011 & 48  & 938  & 257.5 & 926.5  \\
Jan.\ 2012 & 31  & 969  & 39.5  & 966  \\
Feb.\ 2012 & 101 & 1070 & 66  & 1032 \\
Mar.\ 2012 & 71  & 1141 & 86  & 1118 \\
\hline
\end{tabular}
\end{table}
\section{Gridsearch Calibration}
    The model contains four free parameters which require calibration: the within-property spreading rate $\beta_w$, the between-property spreading rate $\beta_b$, the detection threshold $D$, and the initial seed property $s_0$. We test $1,300$ different seed properties located in the Bay of Plenty region and test several different values of the detection threshold $D$, where set $D = \{ 0.2,0.4,0.6,0.8\}$.
    
\begin{figure}[H]
    \centering
    \includegraphics[width=.8\linewidth]{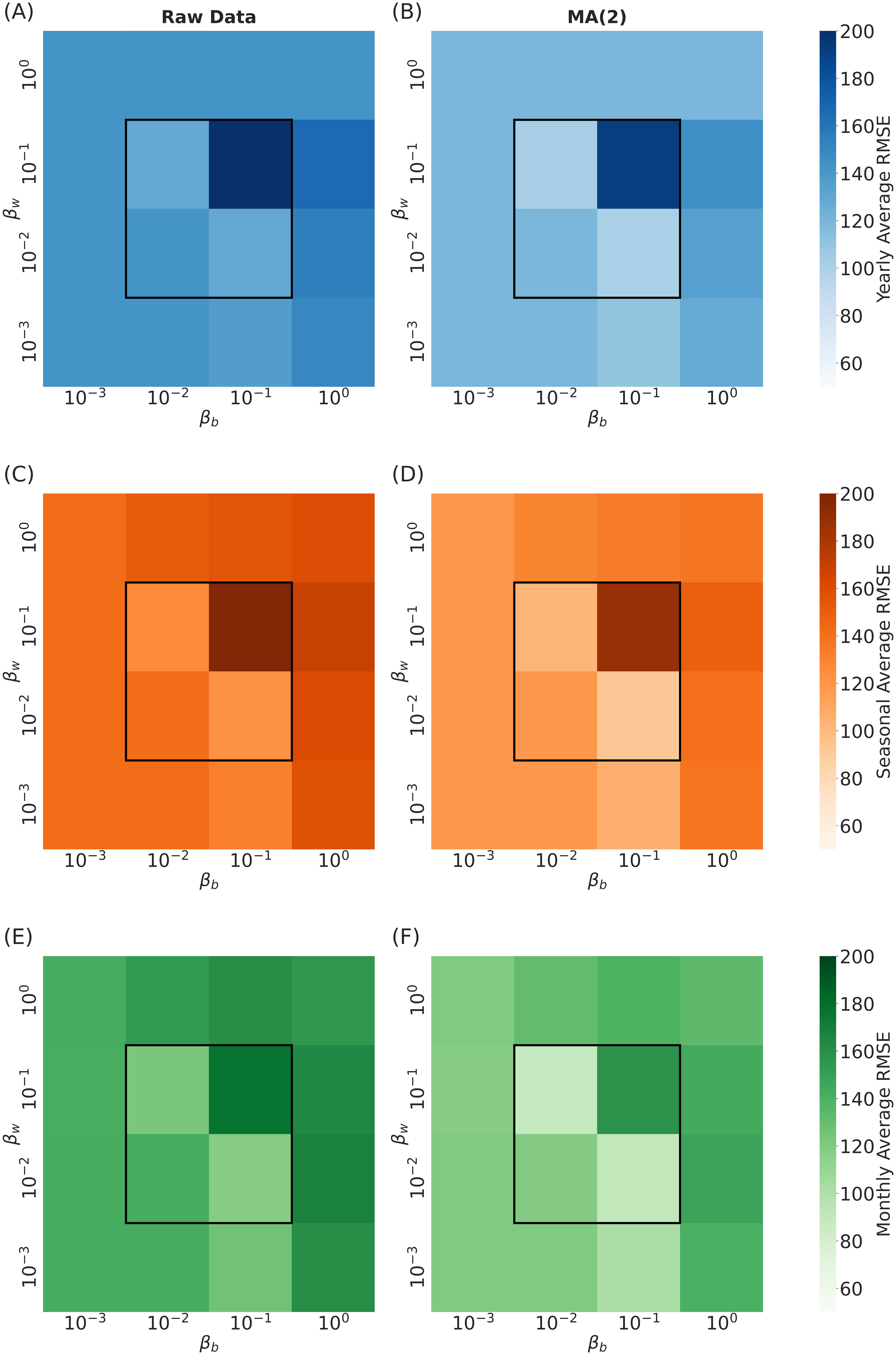}
    \caption{\textbf{Logarithmic gridsearch calibration of $\beta_b,\beta_w$} We present the average RMSE for the various combinations of $(\beta_w,\beta_b)$ where we vary the values for these dynamical parameters from $10^{-3}$ to $10^{0}$ in logarithmic increments.  Each square in the heatmap is the expected RMSE value for the joint distribution of $\beta_w$ and $\beta_b$ over all the possible $s_0$ and $D$ combinations.  The black square within each heatmap represent the values for $\beta_b,\beta_w$ that were searched for in more depth in the linear gridsearch used in the main text. The RMSE values in the first column of heatmaps is calculated with reference to the raw data, while the values in the second column are calculated with respect to the two month moving average (MA(2)) data. In panel (A) we plot the RMSE for the yearly aggregation using the real data , in panel (B) we plot the average RMSE for the yearly network aggregation using the MA(2) data, in panel (C) we plot the average RMSE for the seasonal network aggregation using the raw data. while in panel (D) we plot the average RMSE for the seasonal aggregation using the MA(2) data. In panel(D) and (F) we plot the average RMSE for the monthly aggregation using the raw incidence data and MA(2) incidence data respectively .}
    \label{fig:log-hmap}
\end{figure}

\begin{figure}[H]
    \centering
    \includegraphics[width=.8\linewidth]{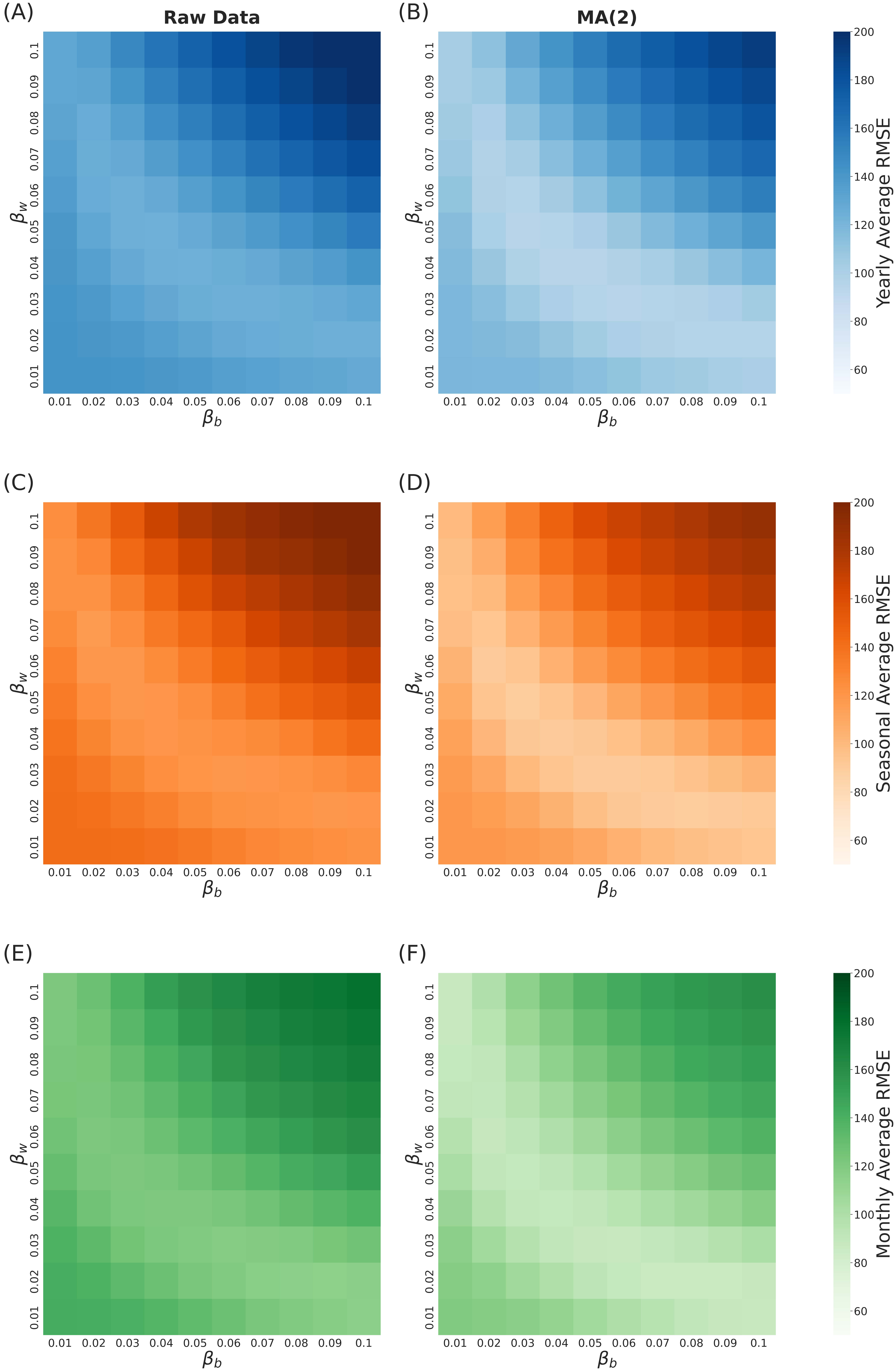}
    \caption{\textbf{Linear gridsearch calibration of $\beta_b,\beta_w$} Similar to Figure ~\ref{fig:log-hmap} we plot the average RMSE for combinations of of $(\beta_w,\beta_b)$, where RMSE values in the first columns of the heatmaps are calculated with reference to the raw data, while the values in the second column are calculated with reference to the MA(2) data. We vary the values for these dynamical parameters from $10^{-2}$ to $10^{-1}$ respectively, in increments of $10^{-1}$.   }
    \label{fig:lin-hmap}
\end{figure}

    To find an appropriate range of values for $\beta_w$ and $\beta_b$,  we first conducted a coarse exploratory gridsearch using logarithmic increments and searched over $[10^{-3},10^0]$. For each pair of  dynamical parameters, $(\beta_w,\beta_b)$,  the average RMSE was calculated using all values of $D$ and $s_0$. The RMSE landscape shown in Figure \ref{fig:log-hmap} indicates that the best parameter configurations are between $[10^{-2}, 10^{-1}]$ for all combinations of $\beta_w,\beta_b$, taking into account different data types and temporal aggregations. Informed by this finding, we performed a finer-grained gridsearch varying $\beta_w,\beta_b$ (Figure ~\ref{fig:lin-hmap}  over the range $[10^{-2}, 10^{-1}]$ using increments of $0.01$ across the set values of $D$ and $s_0$ . 

    \begin{equation}
         |x_c - y_c(\theta)| < \max \{ \epsilon, z* x_c \}
         \label{eq:supp-filter}
    \end{equation}
    
    The average RMSE values shown in both gridsearches (Fig ~\ref{fig:log-hmap} and Fig ~\ref{fig:lin-hmap}) were calculated across the set of configurations $G$, consisting of all possible combinations of $\beta_b,\beta_w,D,s_0$ that were discussed for each gridsearch.  Given $G$, and each configuration entry $\theta$  for all free parameters, we  filtered  configurations using the heuristic defined in the main text to prevent overfitting to the incidence curves.

\begin{figure}[H]
        \centering
        \includegraphics[width=0.5\linewidth]{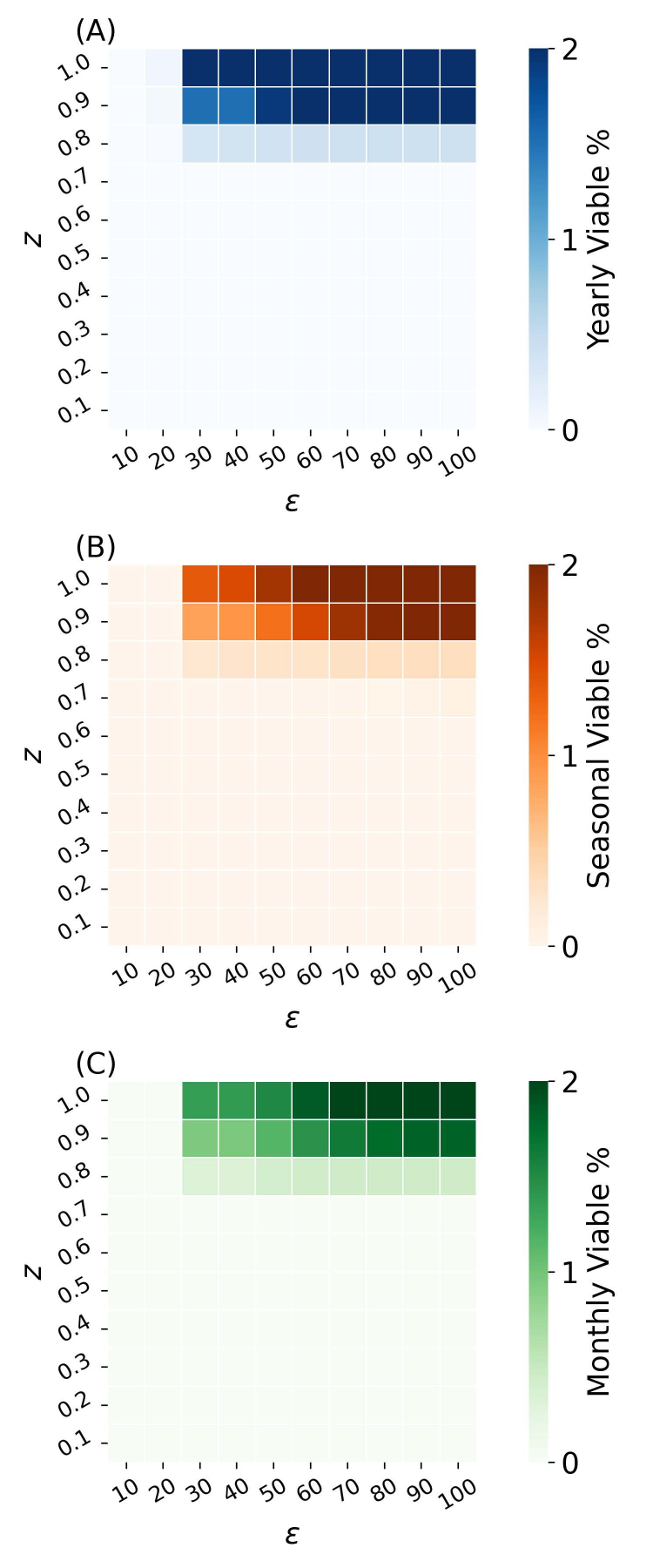}
        \caption{\textbf{Parameter gridsearch for $\epsilon$ and $z$}. The parameter values for $\epsilon$ and $z$ are varied here. We show the results of three different gridsearches, for all aggregations, using the raw data for the incidence curve to determine the . The darker the square in the heatmap, the higher percentage of viable configurations exist for that specified combination of $\epsilon$ and $z$. Panel (A) reports the percentage of viable configurations for the yearly aggregation, (B) and (C) are the same as (A) except for the seasonal and monthly aggregations respectively. }
        \label{fig:find-eps}
    \end{figure}

    We slightly redefine the filtering heuristic in Equation \ref{eq:supp-filter}, using $\epsilon$ and $z$ as the two main variables that need to be set when performing the filtering. $\epsilon$ is utilized as a tolerance value while $z$ is the size of our uncertainty band. In the main text,  $\epsilon = 30$ and $z = 0.8$. A gridsearch was conducted to find the best values for $\epsilon$ and $x$  using the raw data in Figure ~\ref{fig:find-eps}. $\epsilon$ was varied from $10^{1}$ - $10^{2}$ in increments of $10^{1}$ while $z$ was varied from $10^{-1} - 1$ in increments of $10^{-1}$. The percentage of viable configurations that existed by fulfilling the conditions in Equation \ref{eq:supp-filter} was recorded for each combination. From that gridsearch, $\epsilon=30$ and $z=0.8$ are the first two values that produce the smallest percentage of viable configuration among all aggregations. See the main text for details on how we selected the best configuration for each aggregation type. 

% need to show linear gridsearch

    %The set of filtered configurations for each aggregation type $R$ is referred to the viable set $V_{R}$. 

\begin{figure}[H]
    \centering
    \includegraphics[width=\linewidth]{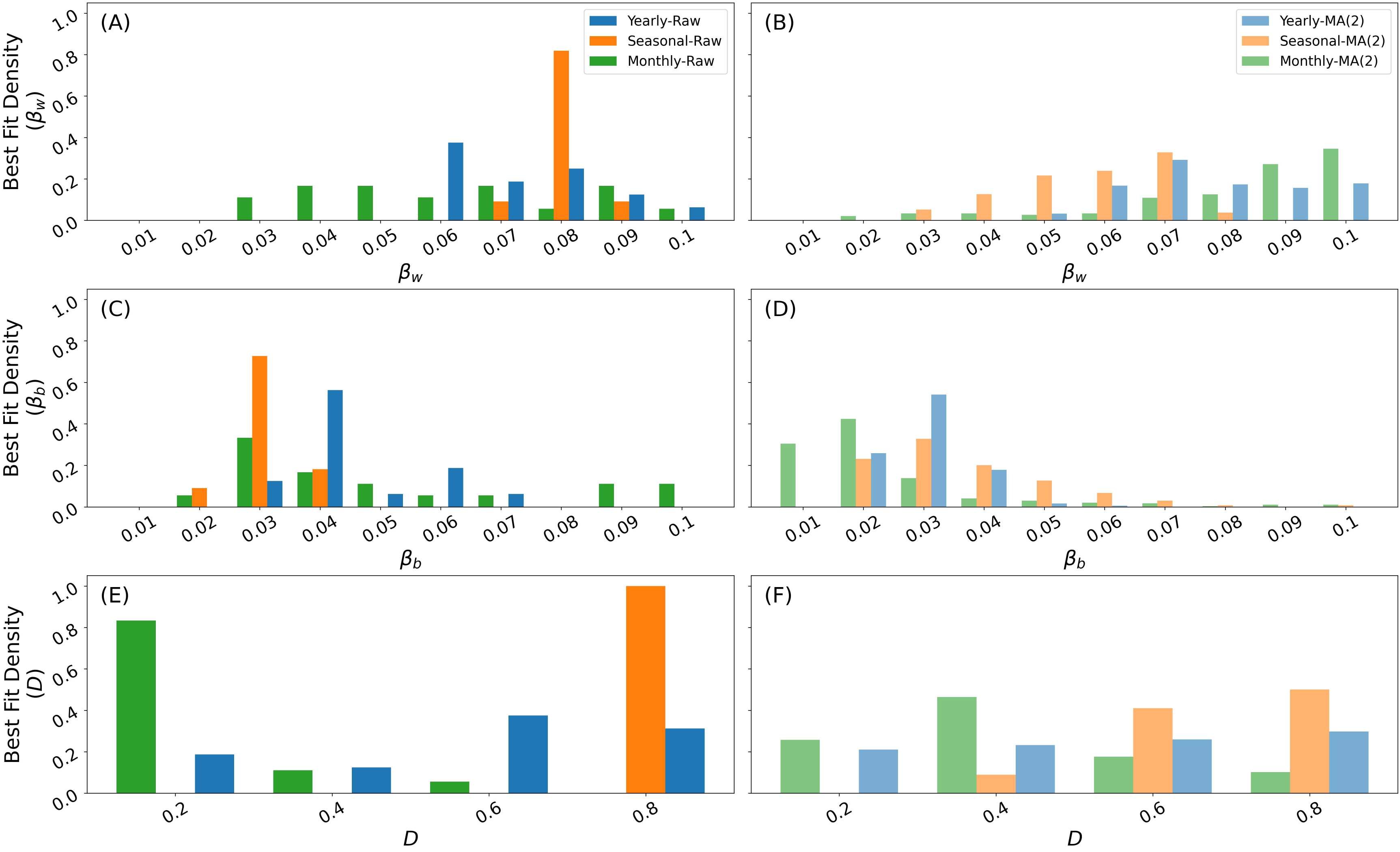}
    \caption{\textbf{Posterior Distribution of Linear Gridsearch} Using the RMSE values generated from the linear gridsearch, seen in Fig. ~\ref{fig:lin-hmap}, we plot the posterior distribution for each parameter (except the seed, due to the number of seeds). These distributions are generated after creating the viable configuration set $V$ for each aggregation and data type (Raw and MA(2) incidence data) and limiting the viable configurations to the configurations that have the smallest $1\%$ of RMSE values among all configurations $V$. The best-fitting configurations are found from the configurations in $V$ that minimize the RMSE best for each configuration and dat type. The first columns RMSE values are calculated with reference to the raw data, while the second columns RMSE values are calculated with reference to the MA(2) data. Three different colors are used in each panel to reference different network aggregations. Blue represents the yearly aggregation, orange represents the seasonal aggregation, and green represents the monthly aggregation. Each row contains the posterior distribution for a different parameter. Panels (A) and (B) plot the distribution for $\beta_w$, panels (C) and (D) plot the distribution for $\beta_b$, and panels (E) and (F) plot the distribution for $D$. The RMSE values in panels (A), (C) and (E) are computed with respect to the raw data while in panels (B),(D), and (E) they are calculated using the MA(2) data. }
    \label{fig:post-dist}
\end{figure}

    Using the viable configurations for each aggregation and data type, we plot the posterior distribution of the parameters in Figure \ref{fig:post-dist}. We observe that there are strongly preferred values for all parameters and aggregations, barring the monthly aggregation for the raw data. For $\beta_w$ and $\beta_b$, specific values are clearly preferred. For the monthly raw aggregation, there is a bimodal distribution. 

\section{Spatial Spread}
\begin{figure}[H]
     \centering
     \includegraphics[width=.75\linewidth]{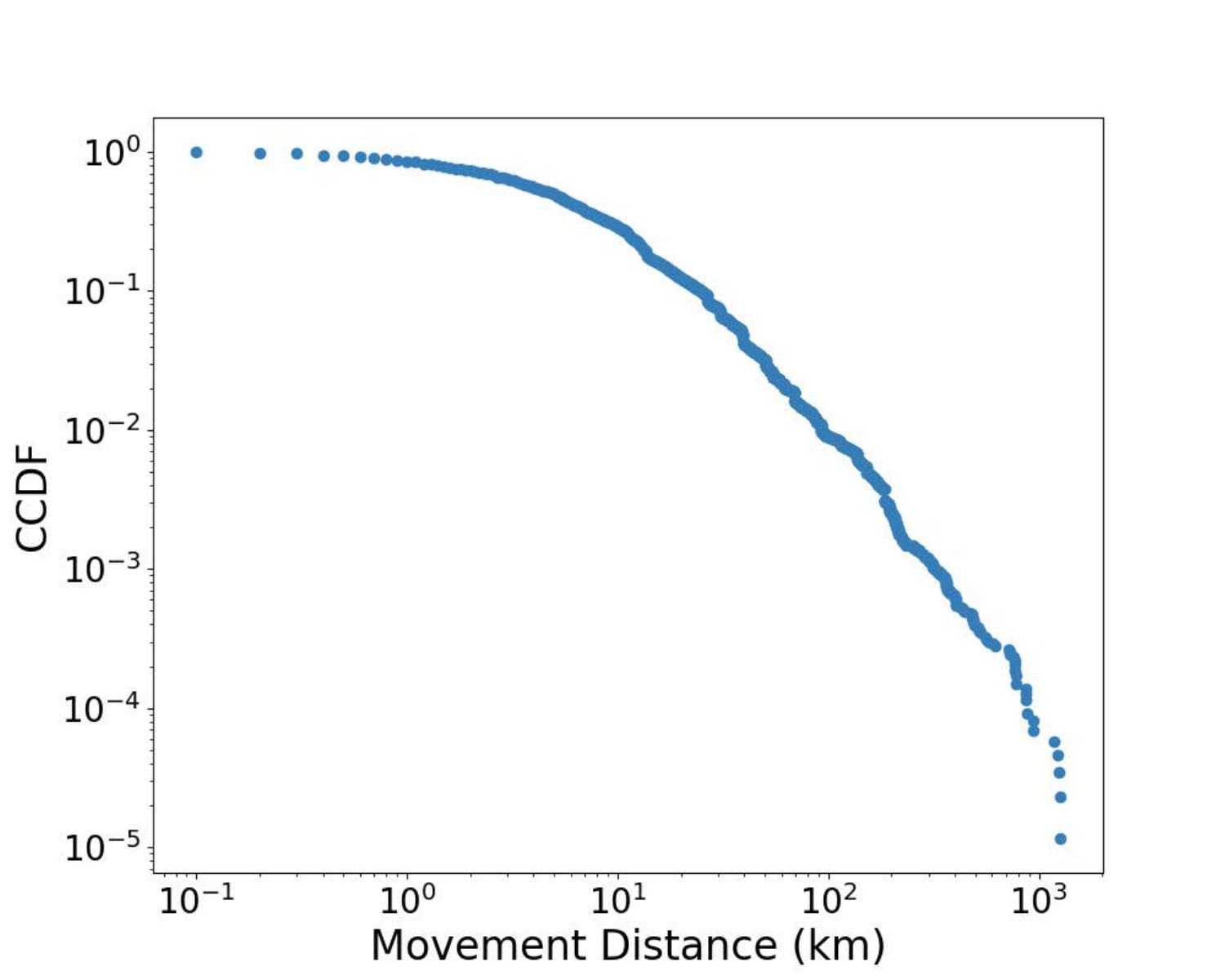}
     \caption{\textbf{CCDF of Movement Length.} }
     \label{fig:ccdf-year}
 \end{figure}

\begin{figure}[H]
    \centering
    \includegraphics[width=.75\linewidth]{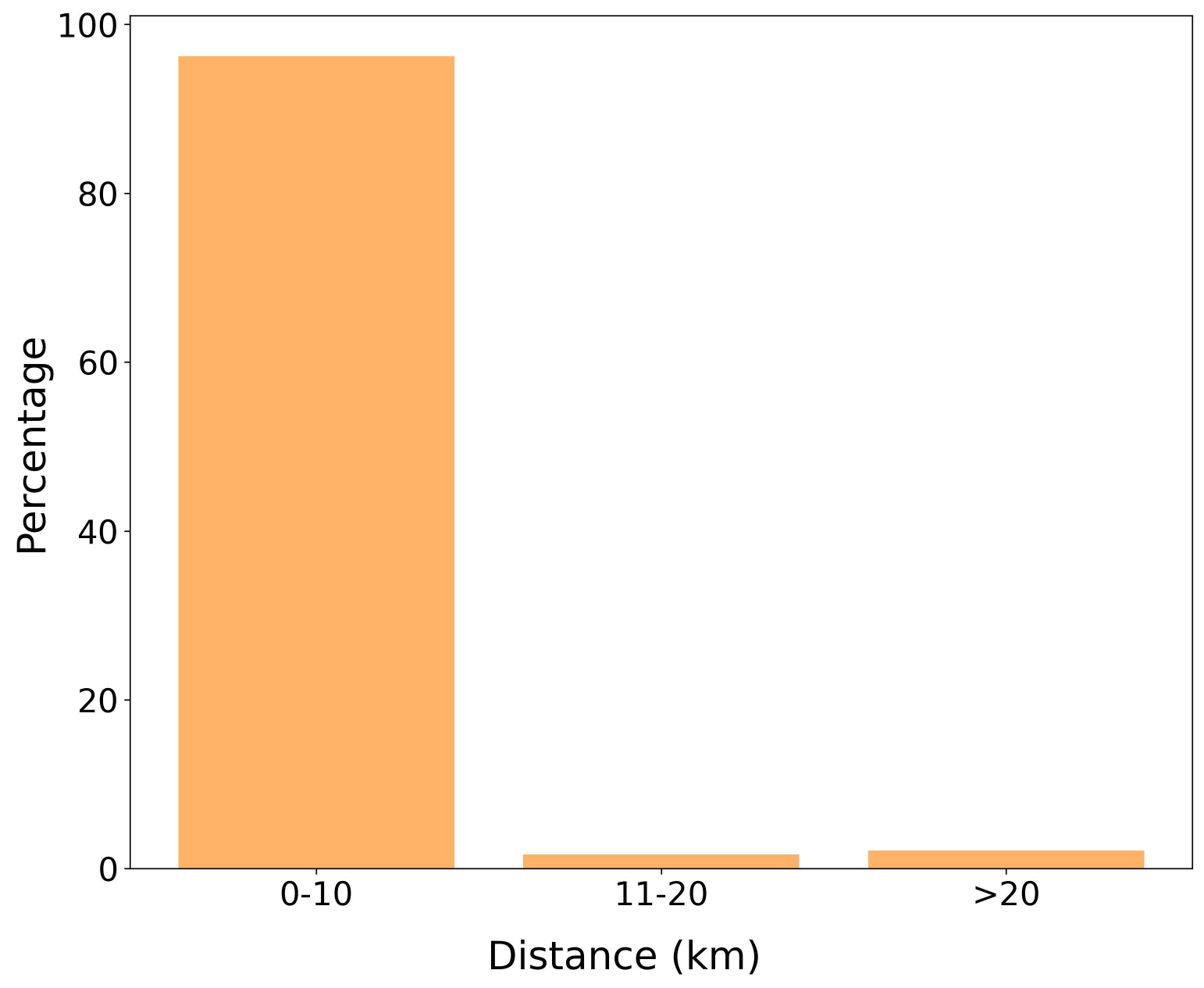}
    \caption{\textbf{Distance of Newly Infected Property to the Closest Infected Property} Once a property becomes infected we measure the distance in kilometers from that property to the closest infected property. An infected property $i$ is defined as one where the level of infection on it exceeds $D$ where $I_i > D$  . We classify this distance using three categories: $0 - 10$ km, $11-20$ km, and $ > 20$ km. Most newly infected properties are within $0-10$ km of a previously infected property, and only $2.7 \%$ of properties are $ > 20$ km away from an infected property. }
    \label{fig:prox-spread}
\end{figure}

\begin{figure}[H]
    \centering
    \includegraphics[width=\linewidth]{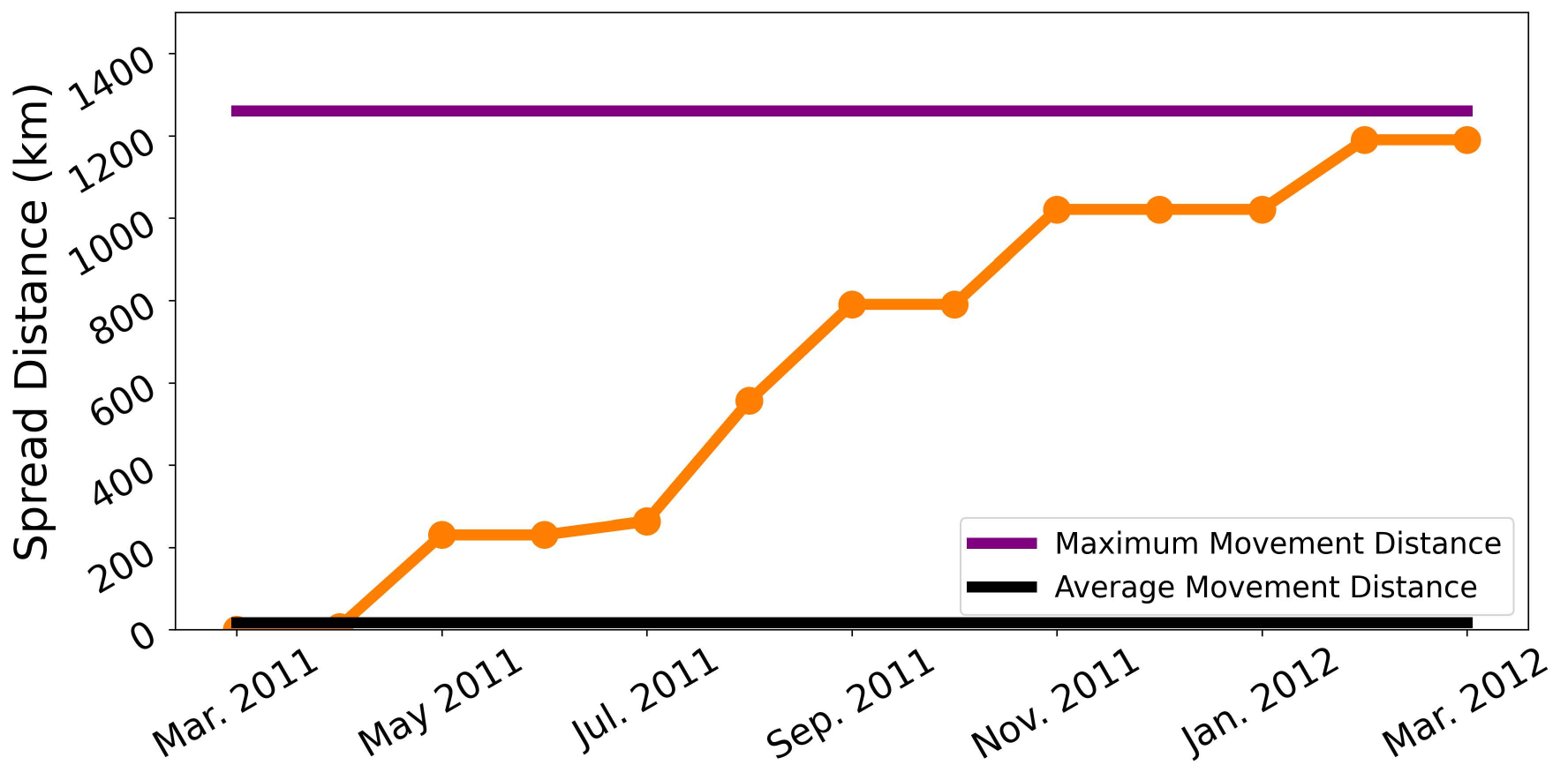}
    \caption{\textbf{Maximum Spread Distance from Seed} Using the calibrated seasonal model, we plot the largest geodesic distance an infected property has from the initially infected seed property $s_0$. We define this distance as the spread distance. The node with the largest distance away from the seed could persist for months, as we observe that for consecutive months at a time the spread distance does not change in value. As reference, we plot the maximum and average movement distance between any two properties as well. By the end of the outbreak, the distance from the seed is very close to the maximum existing movement distance. }
    \label{fig:max-seed}
\end{figure}

The complementary cumulative distribution function (CCDF) of movement length (i.e., distance between origin and destination property) is calculated and shown in Fig.~\ref{fig:ccdf-year}.  
$99.1$\% of movements occur between properties that are between $0-100$ km from each other,  with $71.2$ \% of movements occurring having a distance of $10$ km. From such a movement distribution, we can conclude that most movements in horticulture movement network occur between properties that are relatively close to one another.  This movement structure leads to the results in Figure~\ref{fig:prox-spread} where newly infected properties are infected within $10$ km of another previously infected property. 
While most movements are short-distance, long distance spread from the seed still occurs, as we find in Figure ~\ref{fig:max-seed}, where by the end of the outbreak the epidemic has spread more than $1,000$ km from where it started.

% FIGURE: CCDF 

\newpage
\section{Varying Initial Seed Value}
\begin{figure}[H]
        \centering
        \includegraphics[width=0.8\linewidth]{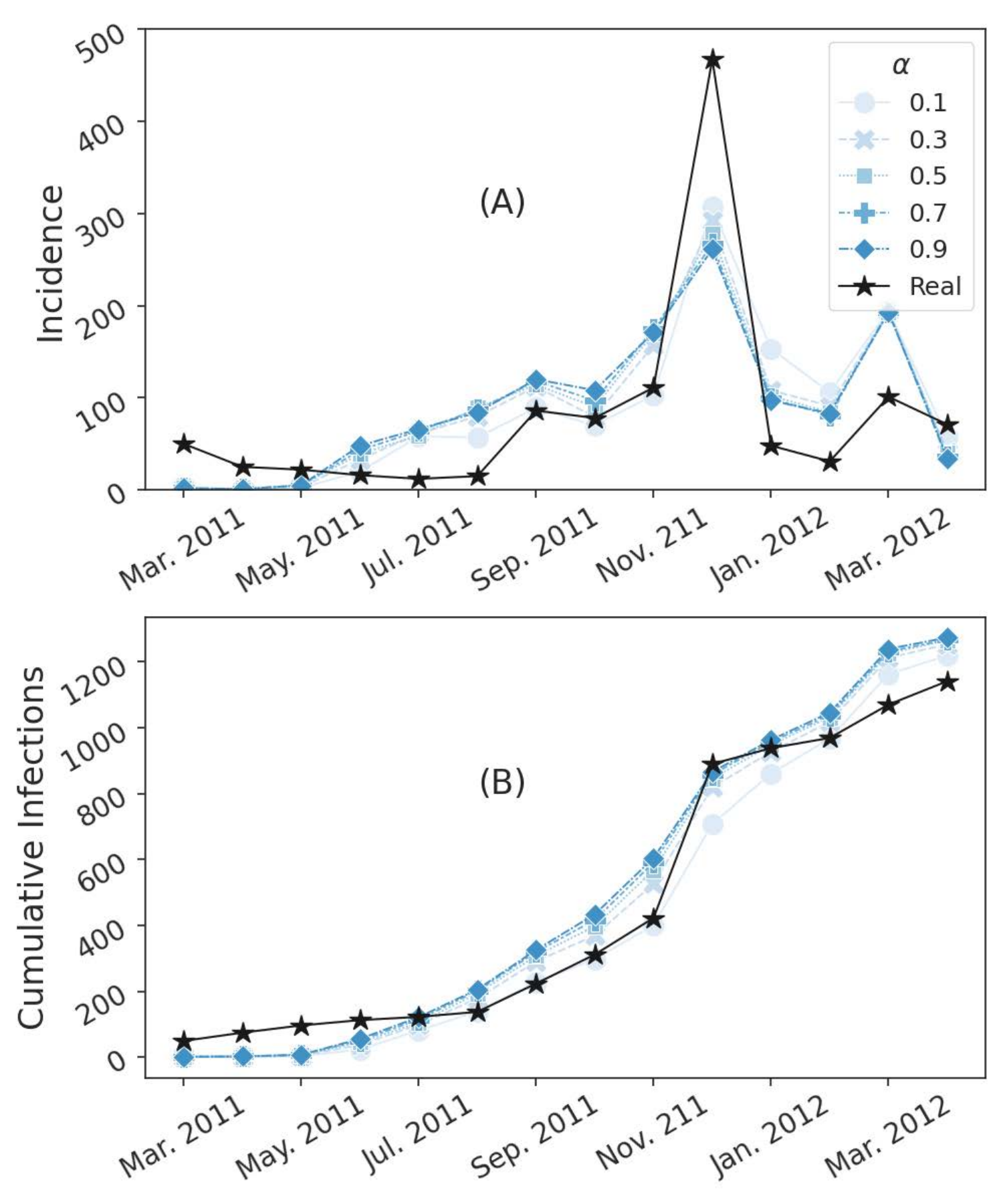}
        \caption{\textbf{Dependence of Model's Outcome on Initial Seed Value} We vary the initial seed value $\alpha$  and plot the effect this has on the incidence curve in panel (A) and on the cumulative total infections shown in panel (B). We set $\alpha = 0.1$ when calibrating the model. Here we vary $\alpha$ from 0.1 to 0.9 in increments of 0.2 in both panels. In panel (A) we plot the real incidence curve as a reference, and in panel (B) the real cumulative infection data is plotted. For all graphs, the seasonal network is used as well as the corresponding parameters of best fit}
        \label{fig:vary-a}
    \end{figure}

We calibrate the model assuming that the seed property's ($s_0$) infection level ($I_{s_0}$) is set to $0.1$. Figure \ref{fig:vary-a} shows the temporal dynamics of the outbreak when the initial amount we infect the seed with, $I_{s_0}$, is varied. We denote $I_{s_0}$ as $\alpha$ for this experiment. We vary $\alpha$ using the calibrated model based on the seasonal network. When calibrating the model $\alpha = 0.1$ , and find that increasing $\alpha$ increases the size of the outbreak while not having a large effect on the incidence dynamics. This result highlights the small effect that $\alpha$ has on outbreak size, and how it does not have a large effect on outbreak dynamics. 
  
%\printbibliography

\bibliography{refs.bib}